\documentclass[11pt]{article}
\usepackage{amsmath}
\usepackage{amssymb}
\usepackage{graphicx,subfigure}
\usepackage{multirow}
\usepackage{booktabs}

\newcommand{\HS}{\ensuremath{\mathcal{H}}}
\newcommand{\IS}[1][N]{\ensuremath{\mathcal{H}_{#1}}}
\newcommand{\PS}{\ensuremath{\mathcal{P}}}
\newcommand{\ket}[1]{\ensuremath{|#1 \rangle}}

\newcommand{\pp}[2][t]{\ensuremath{\frac{\partial #2}{\partial #1}}}

\newcommand{\eps}{\varepsilon}

\newtheorem{prob}{Problem}
\long\def\symbolfootnote[#1]#2{\begingroup%
\def\thefootnote{\fnsymbol{footnote}}\footnote[#1]{#2}\endgroup}

\title{Bohr-Sommerfeld Quantization of Space}

\author{Eugenio~Bianchi${\,}^1$ and Hal~M.~Haggard${\,}^2$\\
{\small ${}^1$Perimeter Institute for Theoretical Physics, 31 Caroline St.N., Waterloo ON, N2J 2Y5, Canada}\\
{\small ${}^2$Department of Physics, University of California,  Berkeley, California USA}} 

\date{August 10, 2012}

\begin{document}

\maketitle

\begin{abstract}
We introduce semiclassical methods into the study of the volume spectrum in loop gravity. The classical system behind a 4-valent spinnetwork node is a Euclidean tetrahedron. We investigate the tetrahedral volume dynamics on phase space and apply Bohr-Sommerfeld quantization to find the volume spectrum. The analysis shows a remarkable quantitative agreement with the volume spectrum computed in loop gravity. Moreover, it provides new geometrical insights into the degeneracy of this spectrum and the maximum and minimum eigenvalues of the volume on intertwiner space.
\end{abstract}

\section{Introduction}

Knowing the classical geometry hiding behind a quantum system provides many insights into its behavior \cite{BrackBhaduriBook}. In this paper we introduce semiclassical methods to investigate the main feature of loop gravity: the prediction of a quantum discreteness of the volume of space \cite{Rovelli:1995,Ashtekar:1995}. We focus on an elementary 4-valent grain of space and derive the semiclassical spectrum of the volume  exploiting knowledge of the classical system associated to it --- a Euclidean tetrahedron. This analysis develops the Bohr-Sommerfeld quantization of space introduced in \cite{Bianchi:2011} and provides a new approximation scheme that can be tested against the numerical results of \cite{DePietri:1996,Thiemann:1998,Carbone:2002,Meissner:2006,Brunnemann:2006,Brunnemann:2008a,Brunnemann:2008b}. Moreover, the semiclassical geometries that arise give new insights into the scaling of the spectrum in a variety of limits.\\

In loop gravity \cite{Rovelli:2004,Thiemann:2007,Ashtekar:2004} quantum states are built on a spin colored graph. The nodes of this graph represent three dimensional grains of space and the links of the graph encode the adjacency of these regions. Traditionally \cite{Rovelli:2010bf,Rovelli:2011eq} the quantum volume of a region of space $R$ is obtained by regularizing and quantizing the classical expression
\begin{equation}
V =  \int_{R} d^{3}x\, \sqrt{h}
\end{equation}
where $h$ is the metric of space. The total volume is obtained by summing the contributions from each node of the spin colored graph contained in the region $R$. Due to the necessity of a regularization scheme there are different proposals for the volume operator at a node. The operator originally proposed by Rovelli and Smolin \cite{Rovelli:1995} is 
\begin{equation}
\hat{V}_{\text{RS}}= \alpha\;\sqrt{\sum_{r<s<t}\big|\vec{E}_r\cdot(\vec{E}_s\times \vec{E}_t)\big|},
\label{eq:RSVol}
\end{equation}
where $\vec{E}_r$ are electric-flux operators associated to each link of the graph impinging on the node, $\alpha>0$ is a multiplicative constant, and the sum over $r<s<t$ spans triples of links at the node. A second operator introduced by Ashtekar and Lewandowski \cite{Ashtekar:1995} is
\begin{equation}
\hat{V}_{\text{AL}}= \alpha\;\sqrt{\Big|\sum_{r<s<t} \eps(v^r,v^s,v^t) \;\vec{E}_r\cdot(\vec{E}_s\times \vec{E}_t)\Big|},
\label{eq:}
\end{equation}
where $v^r$ are the tangents to the links of the spin network graph $\Gamma$ embedded in the space $3$-manifold. The $\eps$ are signs, $\eps(v^r,v^s,v^t)=\pm1$ or $0$ corresponding to whether the triple of tangents is right handed, left handed or planar respectively. When $\hat{V}_{\text{AL}}$ is viewed  as an operator on the intertwiner Hilbert space associated to the node, the tangents $v^r$ and the signs $\eps(v^r,v^s,v^t)$ have to be understood as external fixed data. This residual dependence on the embedding of the spin network graph plays an important role in the definition of Thiemann's regularization of the Hamiltonian constraint \cite{Thiemann:2007}.

Both the Rovelli-Smolin and the Ashtekar-Lewandowski proposals, defined here on the node Hilbert space, admit classical versions: we dequantize the operators $\vec{E}_{l}$ to obtain vectors $\vec{A}_{l}\in \mathbb{R}^3$. This results in two distinct functions on phase space
\begin{equation}
V_{\text{RS}}(\vec{A}_l)\qquad \text{and}\qquad V_{\text{AL}}(\vec{A}_l).
\label{eq:}
\end{equation}

Recently, a third proposal for the volume operator at a node has emerged \cite{Bianchi:2011a}. Motivated by the geometry of the Minkowski theorem, Bianchi, Don\'a and Speziale suggest the promotion of the classical volume of the polyhedron associated to $\{\vec{A}_{l}\}$ to an operator
\begin{equation}
\label{eq:PolVol}
V_{\text{poly}}(\vec{A}_{l}) \qquad \rightarrow \qquad \hat{V}_{\text{poly}}( \vec{E}_{l}).
\end{equation}
The number $N$ of links at the node determines the number of faces of the polyhedron. One advantage of this proposal is that it is closer in structure to the spin foam formulation of the dynamics of loop gravity \cite{Perez:2012wv,Rovelli:2011eq}. \\

In the case of a node with four links, $N=4$, all three of these proposals for the volume operator coincide and match the operator introduced by Barbieri \cite{Barbieri:1998} for the volume of a quantum tetrahedron. The heart of this paper is a study of the semiclassics of this operator. Remarkably, the space of convex polyhedra with fixed face areas has a natural phase space structure. This was first shown by Kapovich and Millson \cite{Kapovich:1996} in a different context (see also \cite{Bianchi:2011a,Baez:1999tk,Conrady:2009px}). Using this kinematics we study the classical volume dynamics and perform a Bohr-Sommerfeld quantization of its spectrum. The semiclassical analysis provides a useful synthesis and simplification of previous results: we find excellent agreement with the studies of the volume operator spectrum \cite{DePietri:1996,Thiemann:1998,Carbone:2002,Meissner:2006,Brunnemann:2006,Brunnemann:2008a,Brunnemann:2008b} and provide a new derivation of many of its features.  Moreover, our analysis provides new geometrical insights on analytic properties of the volume spectrum, such as its degeneracy and the values of the maximum and minimum eigenvalues; for a detailed list of the new results see the conclusions in section \ref{sec:VolConcs}. \\

We briefly recall the central ideas of the semiclassical methods used in this paper. The later development of Bohr's correspondence principle by Sommerfeld and Ehrenfest led to an elegant approximate quantization, now called Bohr-Sommerfeld quantization. The technique begins by finding the dynamical orbits of the system. In our case, the role of the Hamiltonian is played by the volume and we will analyze the volume orbits in phase space.\footnote{More precisely, we use the oriented volume square $Q$ to simplify the analysis.} The Bohr-Sommerfeld quantization condition is then expressed in terms of the action integral $I$ associated to each of these orbits:
\begin{equation}
\label{eq:BSE}
I(E) = \oint p \, dx =  2 \pi \hbar \,\big(n+\frac{1}{2}\big),
\end{equation}
where the integral is over the orbit of energy $E$ and the quantization level is denoted by $n$ . Solving for $E_n$ gives a semiclassical approximation for the volume spectrum.

It is interesting to notice that we are deriving the spectrum of the spatial volume in quantum gravity using methods that are older than quantum mechanics itself. In the `old quantum theory', the quantization condition \eqref{eq:BSE} was based on a physical assumption about robustness of discrete spectra under slow perturbations. Already several systems were known to have discrete equispaced spectra, for instance the harmonic oscillators in Planck's work on the black body spectrum, or in Einstein and Debye's work on specific heats. If the parameters of these systems are slowly modified, the energy spectrum changes but its discreteness is not destroyed. Lorentz and Einstein asked about a classical version of this robustness of spectra: i.e. whether there was a classical quantity that could be preserved under a slow --- adiabatic --- change of the external parameters of such systems? This question led Ehrenfest to identify \emph{adiabatic invariants} as quantities that, on allowed orbits, are realized in multiples of $2\pi \hbar$. The Jacobi action integral $I(E)$  can be shown to be an adiabatic invariant for the system, and provides a natural candidate to extend Planck's quantization of the energy of the harmonic oscillator to other periodic systems, for example the Hydrogen atom. These methods continue to offer new insights into quantum physics by exposing the classical geometries underlying quantum phenomenon in the semiclassical limit.

The structure of the paper is as follows: We begin by summarizing the setup for the node Hilbert spaces $\IS$ for general valency $N$. Next we describe how this space limits to a classical phase space and describe its Poisson structure.  In section \ref{sec:LitRev} we briefly describe what is known about the volume spectrum in loop gravity,  focusing on the case $N=4$ of the quantum tetrahedron. In sections \ref{sec:ClassVol} and \ref{sec:BSTet} we describe the phase space of a classical tetrahedron, study the volume dynamics, and take up its Bohr-Sommerfeld quantization. The spectrum and its degeneracies are reported in sections \ref{sec:BSTet} and \ref{subsec:ReggeSymmetry}. Elliptic functions play a prominent role in this quantization and  section \ref{sec:LimCases} develops tools for analyzing the largest and smallest volume eigenvalues using the properties of elliptic functions. Section \ref{sec:VolConcs} summarizes our results.

\section{Volume operator in loop gravity: the 4-valent node case}
\label{sec:LitRev}\label{sec:QuantTet}

The graph Hilbert space $\HS_{\Gamma}$ of loop gravity can be built out of gauge-invariant Hilbert spaces associated to nodes of the graph. For the entirety of this paper we focus on a single node $n$ and its Hilbert space $\IS$. Let us assume that the node is $N$-valent and call the spins labeling its links $j_{l} \ (l=1, \dots, N)$. To each representation $j_l$ we associate a vector space $\HS_{j_l}$ that carries the action of the $SU(2)$ generators $\vec{J}_{l}$. The standard basis $\ket{j_l m_l}$ is labelled by eigenvalues of the Casimir $J_{l}^{2} = \vec{J}_{l} \cdot \vec{J}_{l}$ and the $z$-component of $\vec{J}_{l}$, $J_{l z}$.  The space $\IS$ is defined as the subspace of the tensor product $\HS_{j_1} \otimes \cdots \otimes \HS_{j_N}$ that is  invariant under global $SU(2)$ transformations (the diagonal action)
\begin{equation}
\mathcal{H}_{N} = \text{Inv} \left( \HS_{j_1} \otimes \cdots \otimes \HS_{j_{N}} \right).
\end{equation}
We call $\IS$ the space of intertwiners and when emphasizing the parametric dependence on $j_1, \dots, j_N$ we write $\IS(j_1, \dots, j_N)$. The diagonal action is generated by the operator $\vec{J}$,
\begin{equation}
\label{eq:sumJs}
\vec{J} = \sum_{l=1}^{N} \vec{J}_{l}.
\end{equation}
States of $\IS$ are called intertwiners and can be expanded in the $\ket{j_{l} m_{l}}$ basis described above, for $\ket{i} \in \IS$,
\begin{equation}
\ket{i}  = \sum_{m\text{'s}} i^{m_1 \cdots m_{N}} \ket{j_1 m_1} \cdots \ket{j_{N} m_{N}}.
\end{equation}
The components $i^{m_{1} \cdots m_{N}}$ transform as a tensor under $SU(2)$ transformations in such a way that the condition 
\begin{equation}
\vec{J} \ket{i} =0
\end{equation}
is satisfied. These are precisely the invariant tensors captured graphically by spin networks. In loop gravity a special role is played by the operators 
\begin{equation}
\vec{E}_{l} = 8 \pi \gamma \ell_P^{2} \vec{J}_{l},
\end{equation}
as illustrated below.\footnote{Here $\ell_P$ is the Planck length and $\gamma$ is the Barbero-Immirzi parameter. They should both be understood as coupling constants of the theory. Throughout the remainder of the paper we will take $\ell_P=\gamma=\hbar=1$.}

Let us now specialize to the case $N=4$, a node with four links. The Hilbert space $\IS[4]$ is the intertwiner space of four representations of $SU(2)$,
\begin{equation}
\IS[4]=\text{Inv}\left(\HS_{j_1} \otimes \HS_{j_2} \otimes \HS_{j_3} \otimes \HS_{j_4}\right).
\label{eq:H=inv}
\end{equation} 
We introduce a basis into this Hilbert space using the recoupling channel $\HS_{j_1}\otimes \HS_{j_2}$ and call these basis states $|k\rangle$. The basis vectors are defined as
\begin{equation}
|k\rangle= \sum_{m_1\cdots m_4} i_k^{m_1 m_2 m_3 m_4}|j_1,m_1\rangle|j_2,m_2\rangle|j_3,m_3\rangle|j_4,m_4\rangle
\label{eq:|k>},
\end{equation}
where the tensor $i_k^{m_1 m_2 m_3 m_4}$ is defined in terms of Wigner $3j$-symbols by
\begin{equation}
i_k^{m_1 m_2 m_3 m_4}=\sqrt{2k+1}\sum_{m=-k}^k(-1)^{k-m}\left(
\begin{array}{ccc}
j_1 & j_2 & k\\
m_1 & m_2 & m
\end{array}
\right) \left(
\begin{array}{ccc}
k & j_3 & j_4\\
-m & m_3 & m_4
\end{array}
\right).
\label{eq:ik}
\end{equation}
The index $k$ ranges from $k_\text{min}$ to $k_\text{max}$ in integer steps with,
\begin{equation}
k_{\text{min}}=\max(|j_1-j_2|,\,|j_3-j_4|)\qquad \text{and}  \qquad k_{\text{max}}=\min(j_1+j_2,\,j_3+j_4).
\label{eq:kminmax}
\end{equation}
The dimension $d$ of the Hilbert space $\IS[4]$ is finite and given by\footnote{This dimension can also be expressed in a symmetrical manner that treats all four $j_r$ on an equal footing,
$d=\text{min}(2 j_1, 2 j_2, 2 j_3, 2 j_4, j_1+j_2+j_3-j_4, j_1+j_2-j_3+j_4, j_1-j_2+j_3+j_4, -j_1+j_2+j_3+j_4) + 1.$
}
\begin{equation}
d=k_{\text{max}}-k_{\text{min}}+1.
\label{eq:dim}
\end{equation}
The states $|k\rangle$ form an orthonormal basis of eigenstates of the operator $\vec{J}_1\cdot\vec{J}_2$. Following Barbieri \cite{Barbieri:1998,Baez:1999tk,Conrady:2009px}, these states can be understood as describing quantum tetrahedra. The operator $\vec{E}_r\cdot\vec{E}_s$ measures the dihedral angle between the faces $r$ and $s$ of the quantum tetrahedron \cite{Major:1999}. 

The operator $\sqrt{\vec{E}_r\cdot \vec{E}_r}$ measures the area of the $r$th face of the quantum tetrahedron and states in $\IS[4]$ are area eigenstates with eigenvalues $8 \pi \gamma \ell_P^2\sqrt{j_r(j_r+1)}$,
\begin{equation}
\sqrt{\vec{E}_r\cdot \vec{E}_r}\;|i\rangle=8 \pi \gamma \ell_P^2\sqrt{j_r(j_r+1)}\;|i\rangle.
\label{eq:}
\end{equation}
The volume operator introduced by Barbieri is 
\begin{equation}
\hat{V}=\frac{\sqrt{2}}{3}\;\sqrt{\big|\vec{E}_1\cdot(\vec{E}_2\times\vec{E}_3)\big|} ,
\label{eq:Vbarbieri}
\end{equation}
and because of the closure relation
\begin{equation}
(\vec{E}_1+\vec{E}_2+\vec{E}_3+\vec{E}_4)|i\rangle=0
\label{eq:closure4}
\end{equation} 
this operator coincides with the Rovelli-Smolin operator for $\alpha=2\sqrt{2}/3$. It also follows from \eqref{eq:closure4} that the Ashtekar-Lewandowski operator on $\IS[4]$ is simply given by $\hat{V}_{AL}=\sqrt{|\sigma|}\hat{V}_{RS}$, where $\sigma$ is a number depending on the Grot-Rovelli class, see \cite{Fairbairn:2004,Grot:1996}, of the link tangents at the node that can attain the values $\sigma=0,\pm 1, \pm 2, \pm 3, \pm 4$.\footnote{The number $\sigma$ is defined as 
\begin{equation}
\sigma(v^1,v^2,v^3,v^4)=\eps(v^1,v^2,v^3)-\eps(v^1,v^2,v^4)+\eps(v^1,v^3,v^4)-\eps(v^2,v^3,v^4).
\label{eq:sigma}
\end{equation}}
Therefore the Ashtekar-Lewandowski volume operator coincides numerically with the operators \eqref{eq:RSVol} and \eqref{eq:PolVol}  when the tangents to the links fall into the class corresponding to $\sigma=\pm 4$ and the constant $\alpha$ is choosen to be $2\sqrt{2}/3$ as before. Otherwise, it is proportional to it.
The volume operator introduced by Barbieri can be understood as a special case of the volume of a quantum polyhedron discussed in \cite{Bianchi:2011a}.

In order to compute the spectrum of the volume operator, it is useful to introduce the operator $\hat{Q}$ defined as
\begin{equation}
\hat{Q}=\frac{2}{9}\;\vec{E}_1\cdot(\vec{E}_2\times\vec{E}_3).
\label{eq:Qhat}
\end{equation}
It represents the square of the oriented volume. The matrix elements of this operator are easily computed and we report them momentarily. 
The eigenstates $|q\rangle$ of the operator $\hat{Q}$,
\begin{equation}
\hat{Q}\,|q\rangle= q\,|q\rangle ,
\label{eq:q}
\end{equation}
are also eigenstates of the volume. The eigenvalues of the volume are simply given by the square-root of the modulus of $q$,
\begin{equation}
\hat{V}\,|q\rangle= \sqrt{|q|}\,|q\rangle .
\label{eq:Vq}
\end{equation}

The matrix elements of the operator $\hat{Q}$ in the basis $|k\rangle$ were originally computed independently by Chakrabarti and then by L\'evy-Leblond and L\'evy-Nahas \cite{Chakrabarti:1964,LevyLeblond:1965}. They are 
\begin{equation}
\hat{Q}=(8\pi \gamma L_P^2)^3\!\sum_{k=k_{\text{min}}+1}^{k_{\text{max}}}\!\!\!2 i \,\frac{ \Delta(k, A_1, A_2)\Delta(k,A_3,A_4)}{\sqrt{k^2-1/4}}\,\Big(|k\rangle\langle k-1|-|k-1\rangle\langle k|\Big) 
\label{eq:sumk}
\end{equation}
here we introduce the shorthand $A_l=j_l+1/2$, indeed in the semiclassical limit the operator $\vec{J}_l$ with Casimir $j_l$ is associated to an angular momentum vector $\vec{A}_{l}$ with magnitude $A_l = j_l+1/2$. In fact, as explained in \cite{Aquilanti:2007}, semiclassical quantization of these angular momenta, which in this case is exact, gives a discrete area spectrum coincident with that of loop gravity.  The function $\Delta(a,b,c)$ returns the area of a triangle with sides of length $(a,b,c)$ and is conveniently expressed in terms of Heron's formula
\begin{equation}
\Delta(a,b,c)=\frac{1}{4}\sqrt{(a+b+c)(a+b-c)(a-b+c)(-a+b+c)}.
\label{eq:Heron}
\end{equation}
Computing the spectrum of $\hat{Q}$ amounts to computing the eigenvalues of a $d\times d$ matrix, where $d$ is the dimension of the Hilbert space given in (\ref{eq:dim}). This can be done numerically and several of our figures compare the eigenevalues calculated in this manner to the results of the Bohr-Sommerfeld quantization, see Sections \ref{sec:BSTet} and \ref{sec:LimCases}.

There are a number of properties of the spectrum of $\hat{Q}$ (and therefore of $\hat{V}$) that can be determined analytically. We list some of them below and refer to L\'evy-Leblond and L\'evy-Nahas \cite{LevyLeblond:1965} for a detailed analysis:
\begin{itemize}
	\item The spectrum of $\hat{Q}$ is non-degenerate: it contains $d$ distinct real eigenvalues. This is a consequence of the fact that the matrix elements of $\hat{Q}$ in the basis $|k\rangle$ determine a $d \times d$ Hermitian matrix of the form 
\begin{equation}
\left(\begin{array}{cccc}
0 & i a_1 & 0 &\cdots \\
-i a_1 & 0& i a_2	&\ddots \\
0 & -i a_2 & 0 &\ddots \\
\vdots & \ddots & \ddots& \ddots
\end{array}\right)
\label{eq:Qmatrix}
\end{equation}
with real coefficients $a_i$. For a derivation of this result see Appendix A. 
	\item The non-vanishing eigenvalues of $\hat{Q}$ come in pairs $\pm q$. A vanishing eigenvalue is present only when the dimension $d$ of the intertwiner space is odd. These two properties are a direct consequence of the structure of the matrix (\ref{eq:Qmatrix}) discussed above. This matrix is $i$ times an antisymmetric matrix and the eigenvalues of antisymmetric matrices have this property. Physically, this is a consequence of the action of parity. 
	

\item From the above we see that the volume spectrum of a fixed node space $\IS$, that is, of an intertwiner space with given $j_1, \dots, j_4$, is twice degenerate,
\begin{equation}
\hat{V} |\pm q\rangle = \sqrt{|q|} |\pm q \rangle. 
\end{equation}
There is a second, exact degeneracy in the volume spectrum that has not been previously noted. Let $s = 1/2(j_1+j_2+j_3+j_4)$ and $j_l^{\prime} = s-j_l$ $(l=1,\dots, 4)$. The volume spectrum of the node spaces $\IS[4](j_1, \dots, j_4)$ and $\IS[4]^{\prime}(j_1^{\prime}, \dots, j_4^{\prime})$ are identical. This is a manifestation of the Regge symmetry, which is briefly discussed in Section \ref{subsec:ReggeSymmetry}. 
\item For given spins $j_1, \dots, j_4$, Brunnemann and Thiemann have estimated the maximum volume eigenvalue using Gershgorin's circle theorem \cite{Brunnemann:2006} and they find that it scales as $v_{\text{max}}\sim j_{\text{max}}^{3/2}$, where $j_{\text{max}}$ is the largest of the four spins $j_l$. 
\item They have also estimated the minimum non-vanishing eigenvalue (volume gap) and find that it scales as $v_{\text{min}}\sim j_{\text{max}}^{1/2}\,$  \cite{Brunnemann:2006,Brunnemann:2008b}.
\end{itemize}
Our semiclassical analysis reproduces all of these results and provides several new insights into their structure. For a parallel list of the semiclassical results see the conclusions. This completes our review and we turn now to the classical geometry of tetrahedra.

\section{Tetrahedral volume on shape space}
\label{sec:ClassVol}

The finite dimensional intertwiner space $\IS$ can be understood as the quantization of a classical phase space
\cite{Bianchi:2011a,Baez:1999tk,Conrady:2009px}. We set up this correspondence in three steps: first the relationship between the algebra of vectors and the geometry of convex polyhedra (Minkowski's theorem) is explained, next we describe how to endow these vectors with Poisson and symplectic structures by interpreting them as angular momenta and, finally, we briefly describe how quantization of these objects leads to $\IS$. The section concludes with the specialization of these results to the case of tetrahedra and a discussion of the classical volume.

 Minkowski's theorem, \cite{Minkowski:1897}, states the following: given $N$ vectors $\vec{A}_{l} \in \mathbb{R}^{3} \ (l=1, \dots, N)$ whose sum is zero
\begin{equation}
\vec{A}_{1} + \cdots +\vec{A}_{N} = 0
\end{equation}
there exists, up to rotations and translations, a unique convex polyhedron with $N$ faces associated to these vectors. Furthermore, the vectors $\vec{A}_l$ can be interpreted as the outward pointing normals to the polyhedron and their magnitudes $A_l= |\vec{A}_l|$ as the face areas. Minkowski's proof is not constructive and so this is strictly an existence and uniqueness theorem. We call the process of building one of these polyhedra (given the area vectors) a Minkowski reconstruction. In the case of the tetrahedron the reconstruction is trivial but for larger $N$ this is a difficult problem \cite{Bianchi:2011a}. 
These convex polyhedra are our semiclassical interpretation of the loop gravity grains of space.

Following Kapovitch and Millson \cite{Kapovich:1996}, we can associate a classical phase space to these polyhedra.  We interpret the partial sums
\begin{equation}
p_{k} \equiv \left| \sum_{l=1}^{k+1} \vec{A}_{l} \right|  \qquad \qquad (k=1, \dots, N-3) 
\end{equation}
as generators of rotations about the $\vec{p}_{k} \equiv \vec{A}_{1} + \cdots + \vec{A}_{k+1}$ axis. This follows naturally from interpreting each of the $\vec{A}_{l}$ vectors as a classical angular momentum. There is a well-known Poisson structure on a single angular momentum (the Lie-Poisson bracket, see \cite{Marsden:1999b}) and this extends to the bracket
\begin{equation}
\{f, g\} = \sum_{l=1}^{N} \vec{A}_{l} \cdot \left( \pp[\vec{A}_{l}]{f} \times \pp[\vec{A}_{l}]{g} \right)
\end{equation}
on $N$ distinct angular momenta, here $f$ and $g$ are arbitrary functions of the $\vec{A}_{l}$. With this Poisson bracket the $p_{k}$ do, in fact, generate rotations about the axis $\vec{p}_k = \vec{A}_1+ \cdots +\vec{A}_{k+1}$. This geometrical interpretation suggests a natural conjugate coordinate, namely the angle of the rotation. Let $q_{k}$ be the angle between the vectors
\begin{equation}
\vec{v}_{k} = \vec{p}_k \times \vec{A}_{k+1} \qquad \text{and} \qquad \vec{w}_{k} = \vec{p}_k  \times \vec{A}_{k+2}.
\end{equation}
It is an easy check to show that
\begin{equation}
\{p_{k}, q_{l} \} = \delta_{k l}.
\end{equation}
The pairs $(p_k, q_k)$ are canonical coordinates for a classical phase space of dimension $2(N-3)$; this dimensionality is further explained below. We call this the space of shapes and denote it by $\PS(A_{1}, \dots, A_{N})$ or more briefly $\PS_{N}$.

The explicit inclusion of the $A_l$ in $\PS(A_1, \dots, A_N)$ highlights a parametric dependence of the space of shapes on the magnitudes $A_l$. This parametric dependence is more natural when viewed from another perspective on $\PS_{N}$, i.e. viewing it as a symplectic reduction of the product of $N$ two spheres, $(S^{2})^{N}$. Upon fixing the magnitude $A_l$, the angular momentum vector $\vec{A}_l$ is restricted to a sphere of radius $A_l$. This sphere is a symplectic leaf of the Poisson manifold described above and the collection of all the spheres $(S^{2})^{N}$ can be endowed with the product symplectic structure. If we now symplectically reduce $(S^{2})^{N}$ by the zero level set of the momentum map
\begin{equation}
\vec{A} = \sum_{l=1}^{N} \vec{A}_{l}
\end{equation}
we once again obtain $\PS_{N}$. This also explains the dimension of the reduced space; because zero is a fixed point of the group action we lose twice as many dimensions as there are components of the momentum map, $\dim{\PS_{N}} = 2N-6 = 2(N-3)$.

Finally, that the quantization of $\PS_{N}$ is the Hilbert space $\IS$ of an $N$-valent node $n$ can be seen as follows: Lift the $N$ angular momentum vectors $\vec{A}_l$  via the Schwinger map to a phase space of $2N$ harmonic oscillators, $\mathbb{C}^{2N}$. These oscillators can be quantized in the standard fashion and once again the Schwinger map can be used, on the quantum side, to reduce to $N$ quantum angular momenta. The constraint  $\vec{A}_{1} + \cdots + \vec{A}_N = 0$ becomes
\begin{equation}
\vec{J}_1 + \cdots + \vec{J}_{N} = 0,
\end{equation}
which is precisely the gauge invariance condition \eqref{eq:sumJs}. 

We turn now to the classical analog of formula \eqref{eq:Vbarbieri}, the volume of a tetrahedron as a function on the shape phase space $\mathcal{P}(A_1,\dots, A_4) \equiv \mathcal{P}_{4}$. This will be the starting point of our Bohr-Sommerfeld analysis in the next section. 


As discussed above the Minkowski theorem guarantees the existence and uniqueness of a tetrahedron associated to any four vectors $\vec{A}_l$, $(l=1,\dots,4)$ that satisfy $\vec{A}_{1} +\dots + \vec{A}_{4} = \vec{0}$. Without loss of generality we will take $A_1 \le A_2 \le A_3 \le A_4$. In terms of these magnitudes, a condition for the existence of a tetrahedron is that $A_1 + A_2 +A_3 \ge A_4$, equality yielding a flat (zero volume) tetrahedron. This is clearly necessary, as there would be no way to satisfy closure if $A_1 + A_2 +A_3 < A_4$ held. It is not difficult to argue that this is also sufficient for there to exist at least one tetrahedron with these face areas (in fact, there are infinitely many). The space of tetrahedra with four fixed face areas $\mathcal{P}(A_1,A_2,A_3,A_4) \equiv \mathcal{P}_4$ is, as we will now argue, a sphere. 

Following the general construction outlined above, the canonical coordinates on $\mathcal{P}_4$ are $p_1 = |\vec{A}_{1} + \vec{A}_{2}|$ and $q_1$, the angle between $\vec{v}_{1} = \vec{A}_1 \times \vec{A}_{2}$ and $\vec{w}_{1} = (\vec{A}_1+\vec{A}_2)\times \vec{A}_{3}$. For the remainder of the paper we adopt the simplified notation $\vec{A} \equiv \vec{A}_{1} +\vec{A}_2$, $A = |\vec{A}_1+\vec{A}_2| = p_1$, and $\phi \equiv q_1$. Recalling that the $\vec{A}_{l}, (l=1,\dots, 4)$ vectors are to be thought of as generators of $SU(2)$ actions we observe that $A$ generates rotations of $\vec{A}_1$ and $\vec{A}_2$ about the $\hat{A}$-axis. This action rotates $\vec{v}_1$ about the, perpendicular, $\hat{A}$-axis while leaving $\vec{w}_1$ fixed and thus increments the angle $\phi$. This is the geometrical content of the Poisson bracket relation $\{A, \phi\}=1$. Because $\vec{A}$ is fixed by this rotation the closure condition, 
\begin{equation}
\label{eq:ClosureRel}
\vec{A}_{1} + \vec{A}_{2}+\vec{A}_{3}+\vec{A}_{4} = \vec{A}+\vec{A}_3+\vec{A}_4=0
\end{equation}
 is also unaffected by such a rotation. Consequently, we have a whole circle of distinct tetrahedra for each value of $A$ with $A_{\text{min}} \le A \le A_{\text{max}}$, where $A_{\text{min}} \equiv \max{\{A_2-A_1,A_4-A_3\}}$ and $A_{\text{max}} \equiv \min{\{A_2+A_1,A_4+A_3\}}$. The collection of these circles over the interval of allowed $A$ values is the slicing of a sphere into lines of latitude over the range of its $z$ diameter. This is made precise by the behavior at the ends of the range of $A$;  either the vectors $\vec{A}_1$ and $\vec{A}_2$ or the vectors $\vec{A}_3$ and $\vec{A}_4$ become colinear, and consequently the four $A$-vectors are coplanar. Such configurations are all equivalent up to overall rotations in $\mathbb{R}^{3}$ and hence under the Kapovich-Millson reduction they correspond to a single point in the reduced space.  As usual, the $\phi$ coordinate becomes ill defined at these poles, here this is because either $\vec{v}_1$ or $\vec{w}_1$ vanishes.

\begin{figure}
\begin{center}
   \includegraphics[height=2.6in]{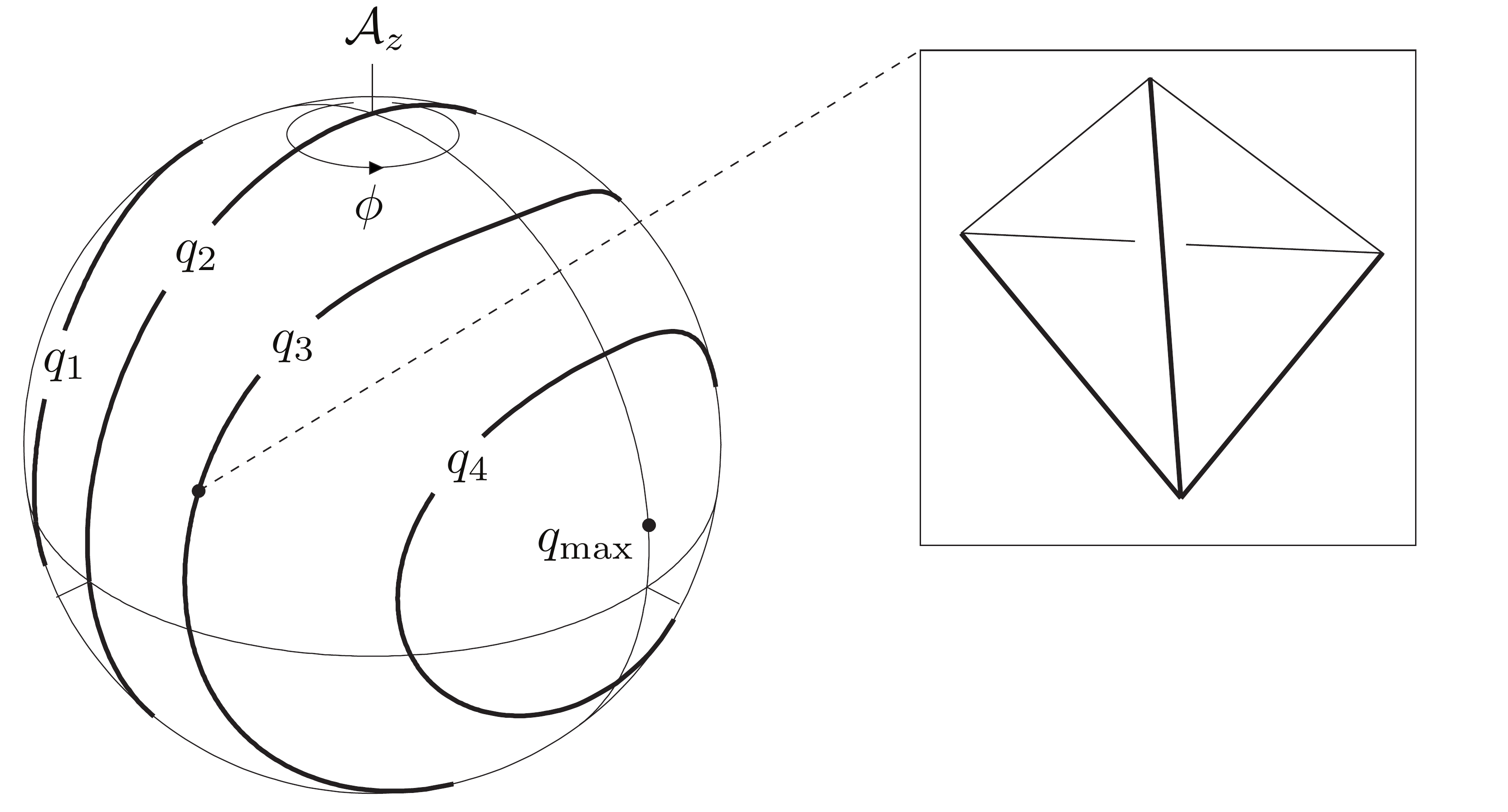}
      \caption[Space of shapes for a tetrahedron]{The space of shapes for a tetrahedron, $\mathcal{P}(\frac{3}{2},\frac{3}{2},\frac{3}{2},\frac{3}{2})$. The darkened contours are quantized level sets of the classical volume squared, $Q$.  One quantized level set is hidden from view. The inset depicts a tetrahedron corresponding to the dot on the $q_3$ level set. The point corresponding to the tetrahedron with the largest possible volume, given the face area constraints, is also marked with a dot. The shape space coordinates are indicated at the top of the sphere. }
      \label{RedVol}
  \end{center}
\end{figure}
The shape space sphere is depicted in Figure \ref{RedVol}. The embedding space is $\mathbb{R}^{3}$ and can be thought of as a copy of the $SU(2)$ Lie algebra associated to $\vec{A}$, we will call this angular momentum space $\mathcal{A}$. Coordinates on $\mathcal{A}$ and the shape space sphere are defined as follows: place the origin of $\mathcal{A}$ at the center of the range of $A$ and choose cartesian coordinates $\mathcal{A}_x$, $\mathcal{A}_y$ and $\mathcal{A}_z$ with $\mathcal{A}_z = A-(A_{\text{max}}+A_{\text{min}})/2$. The radius of the shape space sphere is $R = (A_{\text{max}}-A_{\text{min}})/2$ and the relation $\mathcal{A}_z= R \cos{\theta}$ together with the angle $\phi$ from above define a spherical coordinate system on the sphere. The phase space geometry discussed here is a close analog to that of the symmetry reduced phase space of the $6j$-symbol, see \cite{Aquilanti:2012} for extensive discussion of the $6j$-symbol case.

The classical analog of the volume, \eqref{eq:Vbarbieri}, is
\begin{equation}
V = \frac{\sqrt{2}}{3} \sqrt{| \vec{A}_1 \cdot (\vec{A}_2 \times \vec{A}_3)|},
\end{equation} 
and just as with the quantum theory it will be more straightforward to work with the squared classical volume,
\begin{equation}
Q = \frac{2}{9} \vec{A}_1 \cdot (\vec{A}_2 \times \vec{A}_3).
\end{equation}
A quick vector algebra calculation will convince you that these are indeed the volume and squared volume of a classical tetrahedron. Because $Q$ is a rotational invariant quantity it projects onto the shape phase space and can be thought of as a function of the $A$ and $\phi$ coordinates, $Q(A, \phi)$.\footnote{Because $A$ and $\mathcal{A}_z$ differ by a constant shift we will freely switch between them.} This expression is easily derived by computing $\vec{v}_1 \times \vec{w}_1$; from the definitions of $\vec{v}_1$ and $\vec{w}_1$ one finds
\begin{equation}
\vec{v}_1 \times \vec{w}_1 = \frac{9}{2} Q \vec{A}. 
\end{equation}
Note that the magnitude $|\vec{v}_1| = |\vec{A}_1 \times \vec{A}_2|$ is equal to twice the area $\Delta$ of a triangle with side lengths $A_1, A_2$ and $A$ and, using the closure relation \eqref{eq:ClosureRel}, similarly $|\vec{w}_1|=|\vec{A}_{3}\times \vec{A}_4|$ is twice the area $\bar{\Delta}$ of a triangle with side lengths $A_3, A_4$ and $A$. The definition of $\phi$ as the angle between $\vec{v}_1$ and $\vec{w}_1$ allows us to conclude that the volume squared is
\begin{equation}
\label{edgeVolArea}
Q = \frac{8}{9} \frac{\Delta \bar{\Delta}}{A} \sin{\phi}.
\end{equation}
Calculating the areas $\Delta$ and $\bar{\Delta}$ using Heron's formula \eqref{eq:Heron},
\begin{align}
\Delta 
\label{Triangle1}
 	  &= \frac{1}{4} \sqrt{[(A_1+A_2)^{2}-A^2][A^{2}-(A_1-A_2)^{2}]},\\
\label{Triangle2}
\bar{\Delta} 
 	  &= \frac{1}{4} \sqrt{[(A_3+A_4)^{2}-A^2][A^{2}-(A_3-A_4)^{2}]},
\end{align}
shows that for fixed $A_1, \dots, A_4$, $Q$ is indeed only a function of the coordinates $A$ and $\phi$. The expression \eqref{edgeVolArea} will be the central tool of our Bohr-Sommerfeld quantization. In anticipation of the results of the next section, some quantized level sets of $Q$ are depicted in Figure \ref{RedVol}.

\section{Bohr-Sommerfeld quantization of tetrahedra}
\label{sec:BSTet}
As discussed in the previous section, the phase space of a tetrahedron is two-dimensional  with canonical coordinates $A$, $\phi$,
\begin{equation}
\{A,\phi\}=1.
\end{equation}
The oriented volume square $Q(A,\phi)$, Eq. (\ref{edgeVolArea}), is a function on phase space and generates an Hamiltonian dynamics in a parameter time $\lambda$. The Hamilton equations are given by the familiar formulae
\begin{equation}
\frac{d A}{d\lambda}=\{A,Q\}\quad,\quad \frac{d \phi}{d\lambda}=\{\phi,Q\}.
\end{equation}
Several of the orbits generated by $Q$ have been displayed in Figure \ref{RedVol}. Along these orbits $Q$ is preserved
\begin{equation}
Q(A,\phi)\;=\;q.
\end{equation}
As discussed in the introduction, the Bohr-Sommerfeld quantization condition is expressed in terms of the Jacobi action integral $I$ associated to each of these orbits:
\begin{equation}
\label{eq:BS}
I(q) = \oint A  d\phi =  2 \pi (n+\frac{1}{2}),
\end{equation}
here we have denoted the level value of $Q$ by $q$, the quantization level by $n$, and we take units in which $\hbar = 1$. In the main body of the section we calculate $I$ and impose this quantization condition. This is the usual Bohr-Sommerfeld procedure with $Q$ playing the role of the Hamiltonian $H$ and $q$ the role of the energy $E$.\\

We commence by solving the volume dynamics of $A$. The classical evolution of the tetrahedron with $Q$ taken to be the Hamiltonian turns out to be easiest to calculate in terms of $A^{2}$,
\begin{equation}
\label{AQcomm}
\frac{d (A^{2})}{d \lambda} \equiv \{A^{2}, Q\} = 2 A \{A,Q\} = 2A \frac{\partial Q}{\partial \phi},
\end{equation}
where $\lambda$ is defined to be the variable conjugate to $Q$ and $\phi$ is, as in the previous section, the angle conjugate to $A$. The right hand side can be evaluated by differentiating \eqref{edgeVolArea} with respect to $\phi$,
\begin{equation}
\frac{d (A^{2})}{d \lambda} = \frac{16}{9} \Delta \bar{\Delta} \cos{\phi}.
\end{equation}
Equation \eqref{edgeVolArea} can be used again to eliminate the cosine function,
\begin{equation}
\label{eq:DiffEq}
\frac{d (A^{2})}{d \lambda} = \frac{1}{9} \sqrt{(4 \Delta)^{2} (4 \bar{\Delta})^{2}- (2A)^{2} (9Q)^{2}}.
\end{equation}
The argument of the square root plays an important role in what follows and so we introduce a shorthand for it, 
\begin{equation}
\label{explicitP}
 P(A^{2}, Q^{2}) \equiv (4 \Delta)^{2} (4 \bar{\Delta})^{2}- (2A)^{2} (9Q)^{2}. 
\end{equation}
This is a quartic polynomial in $A^{2}$ and while the general expressions for its roots are complicated they simplify when $Q=0$, and so we further define,
\begin{equation}
\label{eq:PDef}
P(A^{2}, Q^{2})= P(A^2,0)-(2A)^2(9Q)^2 \equiv P_{0}(A^{2})- (2A)^{2} (9Q)^{2}.
\end{equation}
Equations \eqref{Triangle1} and \eqref{Triangle2} yield a factored expression for $P_{0}(A^2)$:
\begin{equation}
\label{explicitP02}
P_{0}(A^{2}) = [A^{2}-(A_1-A_2)^{2}][A^{2}-(A_3-A_4)^{2}][(A_1+A_2)^{2}-A^2][(A_3+A_4)^{2}-A^2].
\end{equation}

Taking the shorthand $x\equiv A^{2}$ we can separate variables in \eqref{eq:DiffEq} and integrate to find,
\begin{equation}
\label{lambda}
\lambda(x) = 9 \int_{r_{2}}^{x} \frac{d\tilde{x}}{\sqrt{(\tilde{x}-r_{1})(\tilde{x}-r_{2})(r_{3}-\tilde{x})(r_{4}-\tilde{x})}} ,
\end{equation}
we assume that the four distinct real roots $(r_1, r_2, r_3, r_4)$ of the quartic $P(x\equiv A^{2},Q^{2})$ are ordered as $r_{1}<r_{2}<r_{3}<r_{4}$. This is an elliptic integral; to bring it to the standard Jacobi form we use a M\"obius transformation that brings the quartic to a conventional one with roots $\pm 1, \pm \frac{1}{\sqrt{m}}$.\footnote{To avoid notational conflicts with the intertwiner eigenstates $| k \rangle$ all elliptic functions are written in terms of the elliptic parameter $m\equiv k^{2}$ instead of the elliptic modulus $k$.} This is possible as long as the cross-ratio of the $r$'s is the same as the cross-ratio of the conventional roots, 
 we use this to set the elliptic parameter $m$. Explicitly the substitution is
\begin{equation}
\label{ellipticMod}
z^{2}  = \frac{(r_{4}-r_{2})(r_{3}-x)}{(r_{3}-r_{2})(r_{4}-x)} \qquad \text{and} \qquad m= \frac{(r_{3}-r_{2})(r_{4}-r_{1})}{(r_{4}-r_{2})(r_{3}-r_{1})}.
\end{equation}
After evaluation of the integral and some algebra this leads to the solution,
\begin{equation}
\label{AsqLam}
x( \lambda ) = A^{2}(\lambda)  = \frac{r_{3}(r_{4}-r_{2})- r_{4} (r_{3}-r_{2}) \text{sn}^{2}\hspace{.1em} ( \frac{\lambda}{9 g} ,m)}{(r_{4}-r_{2})-(r_{3}-r_{2})  \text{sn}^{2}\hspace{.1em} ( \frac{\lambda}{9 g} ,m) } ,
\end{equation}
with 
\begin{equation}
\label{eq:gVal}
g\equiv \frac{2}{ \sqrt{(r_{4}-r_{2})(r_{3}-r_{1})}}.
\end{equation}

This is a complete solution of the dynamics. After the specification of a value for the volume, the quartic $P(x,Q^{2})$ can be solved and the volume evolution of the intermediate coupling $A$ is given by \eqref{AsqLam}. The evolution is periodic and the period can be expressed in terms of the complete elliptic integral of the first kind $K(m)$ by $T=9 g\times 2 K= 18 g K$. The fundamental period of the elliptic functions, $T_{0}=4K$, is halved because they appear squared. For definiteness, in what follows we will assume that the elliptic parameter $m$ is less than one. If this is not the case apply the transformation $ \text{sn}\hspace{.1em} ( u ,m) = \frac{1}{\sqrt{m}}\text{sn}\hspace{.1em} (\sqrt{m} u ,\frac{1}{\sqrt{m}})$ and you will find that the effect on \eqref{AsqLam} and \eqref{eq:gVal} is to switch the roles of $r_{1}$ and $r_{2}$ throughout. 

Before proceeding to the calculation of the action $I$ of a curve $A(\lambda)$, we pause to describe some of the properties of the quartic $P(x, Q^{2})$ that will be useful in this calculation. For the value $Q^{2} = 0$ the quartic simplifies and is given by $P_{0}(x) $. In particular $P_{0}(x)$ can be explicitly factored, see \eqref{explicitP02}, and we will call its four positive real roots $\bar{r}_{1} < \bar{ r}_{2} < \bar{ r}_{3} < \bar{ r}_{4}$. For non-zero real values of the volume $Q$, the roots of the quartic equation $P(x,Q^{2})=0$ are given by the intersections of the line $y=324 Q^{2} x$ with the quartic $y=P(x,0)= P_{0}(x)$. This leads to a few general remarks about the roots $\{r_{1},r_{2}, r_{3},r_{4}\}$ of $P(x,Q^{2})$. The largest and smallest roots are always real and satisfy $0<r_{1}<\bar{r}_{1}=\min \{(A_{1}-A_{2})^2,(A_{3}-A_{4})^{2}\}$ and $r_{4}> \bar{ r}_{4}= \max \{(A_{1}+A_{2})^{2},( A_{3}+A_{4})^{2} \}$. Meanwhile for small enough $Q$ the middle two roots are also real and satisfy, $r_{2}> \bar{ r}_{2}=\max \{(A_{1}-A_{2})^2,(A_{3}-A_{4})^{2}\}$ and $r_{3}< \bar{ r}_{3} =  \min \{(A_{1}+A_{2})^{2},( A_{3}+A_{4})^{2} \}$. As $Q$ grows the middle two roots coalesce and then go off into the complex plane. These observations are summarized graphically in the first two panels of Figure \ref{QuartGrid}. 

The roots of a polynomial coalesce when the polynomial and its first derivative simultaneously vanish, $P(x)=P'(x)=0$. With the notation introduced above $P(x) \equiv P_{0}(x)-324 x Q^{2}$ (we suppress the $Q$ dependence), two roots will coalesce when $P(x)=0$ and $d P/dx = 0 = P_{0}'(x)-324Q^{2}$ or $Q_{\text{coal}} = 1/18 \sqrt{P_{0}'(x)}$ both hold. In section \ref{subsec:ClassicalCases}  we will show that the latter is precisely the condition needed to achieve the maximum volume of a tetrahedron with four fixed face areas,
\begin{equation}
\label{eq:MaxVol}
Q_{\text{max}} = 1/18 \sqrt{P_{0}'(x)}.
\end{equation}
This means that the quartic roots coalesce precisely when the maximum real volume of the tetrahedron is achieved and so for real volumes we need only consider real positive roots. Below we find that the action is given by complete elliptic integrals and it will be useful to be able to assume that the roots that arise in these formulas are real and positive.  

\begin{figure}[htp]
\begin{center}
   \includegraphics[height=3.2in]{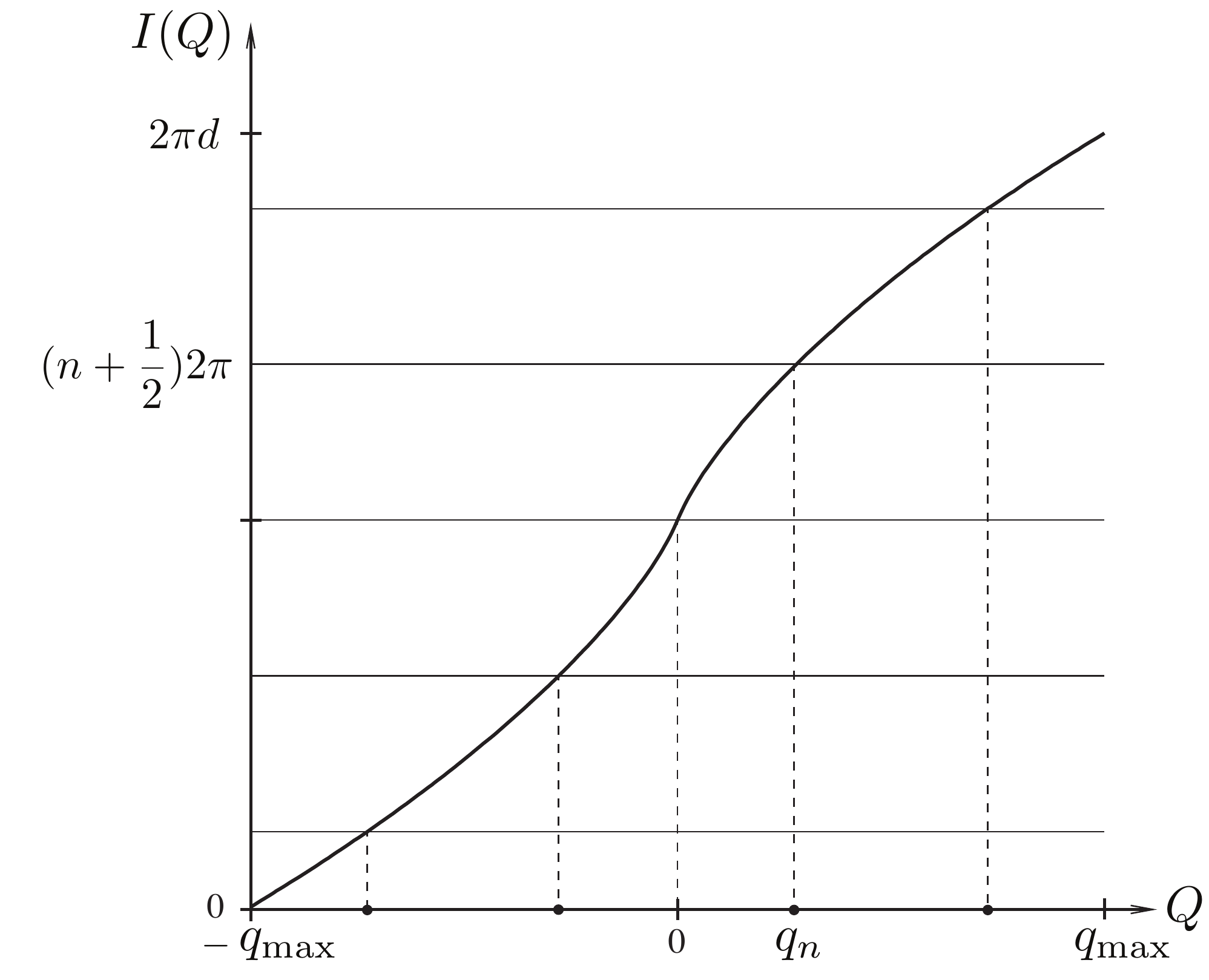}
      \caption[Volume action]{A plot of the action integral $I(Q)$ for the same parameters as Figure \ref{RedVol}. The quantized levels $q_n$ shown satisfy the Bohr-Sommerfeld quantization condition. The corresponding orbits are shown in Figure \ref{RedVol}. }
      \label{fig:ActionPlot}
  \end{center}
\end{figure}

The action integral can now be calculated. The action for an orbit $\gamma$, e.g. one of the curves of Figure \ref{RedVol}, can be re-expressed in terms of the conjugate variable $ \lambda $ to the volume,
\begin{equation}
I = \oint A(\phi) d\phi  = \oint A(\lambda )\frac{d\phi}{d \lambda } d\lambda.
\end{equation}
Once again turning to \eqref{edgeVolArea} and solving for $\phi$ gives, $\phi = \arcsin{( 9 A Q/(8 \Delta \bar{ \Delta }))}$ and differentiating with respect to $\lambda$ yields,
\begin{align}
\begin{aligned}
\frac{d \phi }{d \lambda } &= \frac{ 1}{ \sqrt{ 1-(9 A Q/8 \Delta \bar{ \Delta })^{2}}} \left( \frac{9 Q }{8 \Delta \bar{ \Delta }}  \frac{dA}{d \lambda }- \frac{9 A Q }{8 (\Delta \bar{ \Delta })^{2}} \frac{d (\Delta \bar{ \Delta })}{d \lambda }  \right)\\
				&= \left( \frac{ Q}{A} - \frac{ Q}{ \Delta \bar{ \Delta }} \frac{d ( \Delta \bar{ \Delta })}{d A }  \right).
\end{aligned}
\end{align}
Returning to the expressions for $\Delta$ and $\bar{\Delta}$ (\eqref{Triangle1} and \eqref{Triangle2}) one can calculate $d(\Delta \bar{\Delta})/dA$ and obtain,
 \begin{align}
\oint A \frac{d \phi }{d \lambda } d\lambda = \oint Q d\lambda - \oint Q  &\left( \frac{ A^{2}}{ (A^{2} -(A_{1} +A_{2})^{2})} + \frac{ A^{2}}{ (A^{2} -(A_{1} -A_{2})^{2})}  \right. \nonumber \\
&\left. + \frac{ A^{2}}{ (A^{2} -(A_{3} +A_{4})^{2})} + \frac{ A^{2}}{ (A^{2} -(A_{3} -A_{4})^{2})}  \right)d\lambda.
 \end{align}
Because $Q$ is a constant along the orbit, the first integral simply gives the period of the elliptic function,
\begin{equation}
\oint Q d\lambda =  18 g Q K(m) ,
\end{equation}
where again the elliptic parameter $m$ is given by \eqref{ellipticMod}. The remaining integrals are all of the same type, we introduce a parameter $\bar{r}_{i}$ which takes the values of the roots at zero volume, $(A_{1}-A_{2})^{2}, (A_{3}-A_{4})^{2},(A_{1}+A_{2})^{2},(A_{3}+A_{4})^{2}$, respectively for $i=1, \dots, 4$, and find,
\begin{equation}
\oint Q   \frac{ A^{2}}{ (A^{2} -\bar{r}_{i})} d\lambda = Q \oint  \frac{r_{3}(r_{4}-r_{2})- r_{4} (r_{3}-r_{2}) \text{sn}^{2}\hspace{.1em} ( \frac{\lambda}{9 g} ,m)}{(r_{3}-\bar{r}_{i})(r_{4}-r_{2})-(r_{4}-\bar{r}_{i})(r_{3}-r_{2})  \text{sn}^{2}\hspace{.1em} ( \frac{\lambda}{9 g} ,m) } d\lambda.
\end{equation}
These integrals are more simply expressed in terms of  $u \equiv \lambda/9 g$ and the integration over a complete period is over the interval $u \in [0, 2 K]$. We find,
\begin{align}
\begin{aligned}
\oint Q   \frac{ A^{2}}{ (A^{2} -\bar{r}_{i})} d\lambda &= Q \frac{18 g r_{3}}{(r_{3} - \bar{r}_{i})}  &  &\int_{0}^{K} \frac{1}{1 - \alpha_{i}^{2} \text{sn}^{2}\hspace{.1em} (u,m)} du\\
&&& - Q \frac{18 g r_{4} \alpha_{i}^{2} }{(r_{4} - \bar{r}_{i})} \int_{0}^{K} \frac{ \text{sn}^{2}\hspace{.1em} (u,m)}{1- \alpha_{i}^{2} \text{sn}^{2}\hspace{.1em} (u,m)} du,
\end{aligned}
\end{align}
where
\begin{equation}
\alpha_{i}^{2} = \frac{(r_{4}-\bar{r}_i)(r_{3}-r_{2})}{(r_{3}-\bar{r}_i)(r_{4}-r_{2})}.
\end{equation}
Collecting all four of these integrals we have,
\begin{align}
\label{finalAction}
\begin{aligned}
I &= 18g Q\bigg( K - \sum_{i=1}^{4} && \bigg\{ \frac{r_{3}}{(r_{3}-\bar{r}_{i})}  \int_{0}^{K}   \frac{1}{1 - \alpha_{i}^{2} \text{sn}^{2}\hspace{.1em} (u,m)} du \\
&  &&-\left. \left.\frac{r_{4} \alpha_{i}^{2} }{(r_{4}-\bar{r}_{i})} \int_{0}^{K}\frac{ \text{sn}^{2}\hspace{.1em} (u,m)}{1- \alpha_{i}^{2} \text{sn}^{2}\hspace{.1em} (u,m)} du \right\} \right),
\end{aligned}
\end{align}
which can be evaluated in terms of complete elliptic integrals yielding,
\begin{equation}
\label{finalAction2}
I= 18g Q\left(\left[1-\sum_{i=1}^{4} \frac{r_{4}  }{(r_{4}-\bar{r}_{i})}\right]K(m)-\sum_{i=1}^{4}  \frac{\bar{r}_{i}(r_{4}-r_{3})}{(r_{4}-\bar{r}_{i})(r_{3}-\bar{r}_{i})}\Pi(\alpha_{i}^{2}, m)\right),
\end{equation}
where $\Pi(\alpha_i^{2}, m)$ is the complete elliptic integral of the third kind. To highlight the structure of this result we condense the dependencies on the roots into two coefficients,
\begin{equation}
a \equiv 18 g \left[ 1-\sum_{i=1}^{4} \frac{r_{4}  }{(r_{4}-\bar{r}_{i})} \right] \quad \text{and} \quad b_i \equiv  \frac{ 18 g \bar{r}_{i}(r_{4}-r_{3})}{(r_{4}-\bar{r}_{i})(r_{3}-\bar{r}_{i})} \quad (i=1, \dots, 4).
\end{equation}
This allows us to write $I$ in the more compact form,
\begin{equation}
\label{eq:CompactAction}
I= \left(a K(m)-\sum_{i=1}^{4} b_i \Pi(\alpha_{i}^{2}, m)\right) Q.
\end{equation}
This is our main result. Figure \ref{fig:ActionPlot} displays a plot of this function for the same parameters used in Figure \ref{RedVol}. 

In spite of the complex manipulations used to obtain the expression \eqref{eq:CompactAction} its interpretation is simple: Applying Stoke's theorem to the action $I = \oint A d\phi$ of an orbit $\gamma$,  we can interpret this integral as the symplectic area contained within the orbit. The symplectic form on shape space is determined by the Poisson bracket relation $\{A, \phi \}=1$ and is $\omega = dA \wedge d\phi$ so that
\begin{equation}
I = \oint A(\phi) d\phi = \int \omega = \int dA \wedge d\phi .
\end{equation}
 We can just as well work with $d\mathcal{A}_z \wedge d\phi = R d\cos{\theta} \wedge d\phi$ because $A$ and $\mathcal{A}_{z}$ differ by a constant and so this symplectic area only differs from the solid angle on the sphere by a normalization factor $R$.

\begin{figure*}[t]
   \includegraphics[height=220pt]{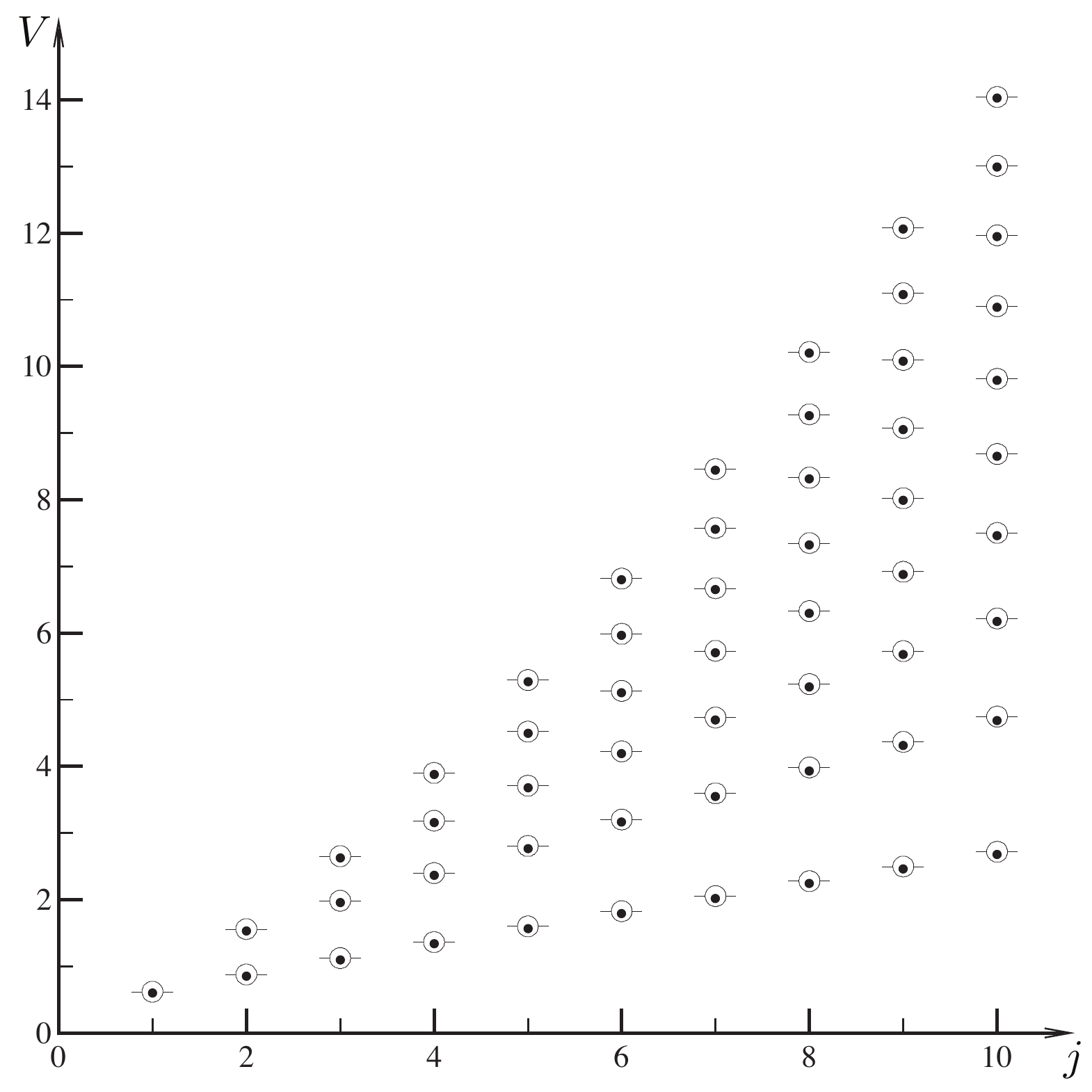} \hspace{25pt}
   \includegraphics[height=220pt]{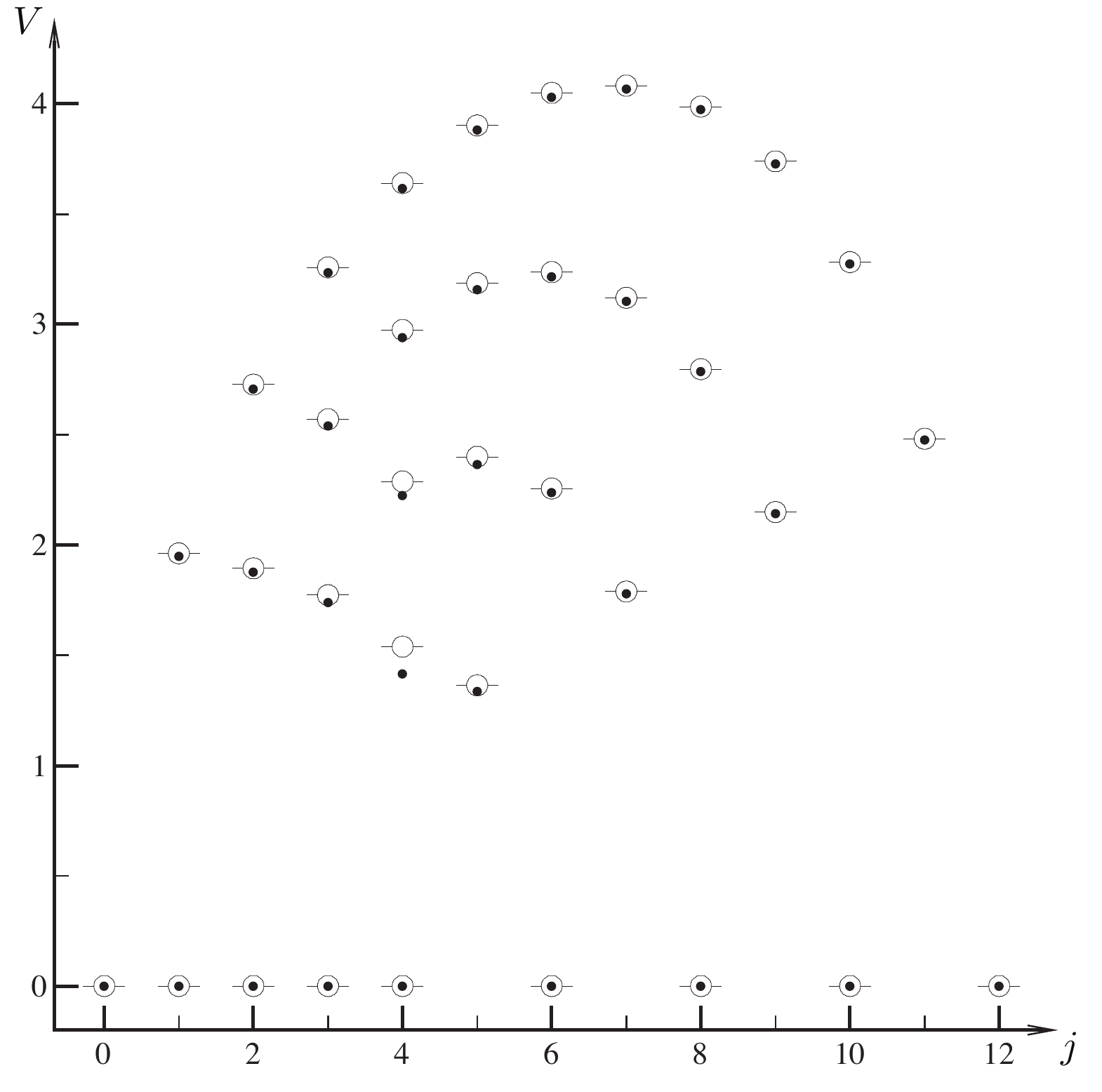} 
   \caption{A comparison of the Bohr-Sommerfeld and loop gravity Volume spectra. On the left: configuration with spins $\{j,\,j,\,j,\,j+1\}$. On the right: configuration with spins $\{4,\,4,\,4,\,j\}$ and $j$ varying in its allowed range. The Bohr-Sommerfeld values of the volume of a tetrahedron are represented as \emph{dots}, the eigenvalues of the loop-gravity volume operator as \emph{circles}. Recall that the spins and areas are related by $A_{l} = j_l+1/2$.}
    \label{fig:Comparison}
\end{figure*}

Although this action has a closed analytic form, finding the Bohr-Sommerfeld spectrum requires a numerical inversion. Recall that the strategy is to find the volumes for which the corresponding orbits capture $(n+1/2)2\pi$ worth of area on the sphere. The analytic expression \eqref{finalAction} has a complicated dependence on the volume; it appears explicitly as an overall multiplicative factor but also implicitly through the elliptic function's dependence on the roots $r_1, \dots, r_4$, all of which depend on $Q$. Instead of trying to invert the action analytically we have calculated its value for several hundred points in the range of classically allowed volumes, interpolated between these values and numerically found the volumes for which the Bohr-Sommerfeld condition is satisfied, this process is illustrated schematically in Figure \ref{fig:ActionPlot}. The results of this analysis are presented for two examples in Figure \ref{fig:Comparison} along with the numerical diagonalization of the volume matrix elements discussed in section \ref{sec:QuantTet}. Recall that $V= \sqrt{|Q|}$ and that the eigenvalues satisfy the same relationship $v= \sqrt{|q|}$ (see \eqref{eq:Vq}). \\

The Bohr-Sommerfeld approximations developed above effectively reproduce the volume spectrum of loop gravity, see Figure \ref{fig:Comparison}. In a manner that is characteristic of this type of semiclassical approximation, the agreement rapidly improves with increasing quantum numbers, in our case increasing $j_l$. This is illustrated in Appendix \ref{app:tables}, which tabulates the Bohr-Sommerfeld and loop gravity volume eigenvalues for a variety of spins. Section \ref{sec:LimCases} contains additional comparisons of the numerical and Bohr-Sommerfeld results. The reduced quality of the approximation for the case where all the spins $j_r$ are equal, exhibited in the right plot of Figure \ref{fig:Comparison} at $j=4$, is also discussed.

\section{Degeneracy of the volume spectrum and the Regge symmetries}
\label{subsec:ReggeSymmetry}

Besides accurately reproducing the frequency of many emission lines in atomic spectra, the old Bohr-Sommerfeld quantization is also able to predict the intensity of such lines, that is, the degeneracy of the energy levels. We now apply these ideas to the case of the volume spectrum.

The volume dynamics generates closed orbits in phase space. In fact, for every allowed non-zero value of the volume there are two closed orbits satisfying $V(A,\phi)=v$. The two orbits are sent one into the other by a parity transformation of the tetrahedron and while the oriented volume square $Q(A,\phi)=q$ is odd under parity, the volume eigenvalue $v=\sqrt{|q|}$ is invariant. Thus the two classical orbits with the same volume give rise to twice degenerate volume levels. The orbit with vanishing volume is invariant under parity and so, $v=0$, when it is an allowed quantum level, is non-degenerate. This is consistent with the following observations about the classical phase space: The total symplectic area of our phase space is finite. The Bohr-Sommerfeld condition implies that this area is divided up into a finite, integer number of Planck cells,
\begin{equation}
\int dA\wedge d\phi\;=\;2\pi\,d,
\end{equation}
here the integer $d$ is also the dimension of the Hilbert space, \eqref{eq:dim}. When $d$ is odd, the Bohr-Sommerfeld condition predicts that there are $d=2n+1$ levels: $n$ doublets of non-vanishing volume and one singlet of vanishing volume. Figure \ref{fig:ActionPlot} provides an example of this case. \\

There is a second more subtle degeneracy: the tabulation of Appendix \ref{app:tables} exposes a degeneracy in the volume spectra of distinct intertwiner spaces. The clearest example of this in our tables, is for the spins $(j_1,j_2,j_3,j_4)=(6,6,6,7)$ and $(j_1^{\prime},j_2^{\prime},j_3^{\prime},j_4^{\prime})= (\frac{11}{2},\frac{13}{2},\frac{13}{2},\frac{13}{2})$; the volume spectra of these two  intertwiner spaces agree exactly. 

This degeneracy can be understood in terms of a classical geometric version of the well known Regge symmetries of the $6j$-symbol \cite{Regge:1959}. Let $s = 1/2(j_1+j_2+j_3+j_4)$, if the spins of a $6j$-symbol,
\begin{equation}
\begin{Bmatrix}
j_1 & j_2 & j_{12}\\
j_3 & j_4 & j_{23},
\end{Bmatrix}
\end{equation}
are transformed to $j_l^{\prime} = s-j_l,$ $(l=1,\dots, 4)$ and $j_{12}^{\prime} = j_{12}, \ j_{23}^{\prime} = j_{23}$, then Regge showed that,
\begin{equation}
\begin{Bmatrix}
j_1 & j_2 & j_{12}\\
j_3 & j_4 & j_{23}
\end{Bmatrix}
=
\begin{Bmatrix}
j_1^{\prime} & j_2^{\prime} & j_{12}^{\prime}\\
j_3^{\prime} & j_4^{\prime} & j_{23}^{\prime}
\end{Bmatrix}.
\end{equation}
Roberts explains that this symmetry can be understood geometrically as a scissors congruence of the relevant tetrahedra \cite{Roberts:1999}. Two polyhedra are scissors congruent if the first polyhedron can be sliced into finitely many polyhedral pieces and then reassembled into the second polyhedron. Evidently, scissors congruence preserves volumes. 

In the present context this symmetry completely explains the degeneracy of the volume spectra; besides the Minkowski tetrahedron having the vectors $\vec{A}_{l}$ as face normals, we prove this using an auxiliary tetrahedron having edge vectors $\vec{A}_{l}$. This auxiliary tetrahedron has volume $\frac{9}{2\times 3!}Q = \frac{1}{3!}\vec{A}_1 \cdot (\vec{A}_2 \times \vec{A}_3)$. Now, performing a Regge transformation we obtain a second auxiliary tetrahedron that is scissor congruent to the first one. The two auxiliary tetrahedra have the same volume, $Q^{\prime}= Q$, and this implies that the volumes of the corresponding Minkowski tetrahedra are also equal $V^{\prime} = \sqrt{|Q^{\prime}|} = \sqrt{|Q|} = V$. Moreover, for every auxiliary tetrahedron in the phase space $\mathcal{P}_4$ we have a scissor congruent tetrahedron in the Regge transformed phase space $\mathcal{P}'_4$. As a result the action integrals also coincide, $I(q)=I'(q)$, and the degeneracy of the volume spectrum is explained.

The Regge degeneracy of the spectrum leads to a conjecture: We have shown above that two Minkowski tetrahedra related by a Regge symmetry have the same volume. Are they also scissor congruent? We conjecture that they are. The proof of this conjecture hinges on showing that these two tetrahedra have the same Dehn or Hadwiger invariants, again see \cite{Roberts:1999}. We leave the investigation of this conjecture open for future work.

\section{Limiting cases: largest and smallest volumes}
\label{sec:LimCases}

\subsection{Classical analysis of limiting cases}
\label{subsec:ClassicalCases}

Before proceeding to a Bohr-Sommerfeld analysis of the limiting values of the volume spectrum, we investigate the classical extrema of $|Q|$. At the classical level, the minimum of $|Q|$ is always zero: Open the angle $\phi$ until the plane spanned by $\vec{A}_1$ and $\vec{A}_{2}$ coincides with the plane spanned by $\vec{A}_3$ and $\vec{A}_{4}$, then $Q=0$. This, however, does not lead to the conclusion that the minimum of $V$ is always zero. This is because the Minkowki theorem does not hold for planar configurations of the $\vec{A}_l$, instead it is generically singular for these configurations.  Certainly $Q=0$ implies that $V=0$, however, the issue is the correspondence between a planar set of $\vec{A}_l$ and such a flat tetrahedron. Geometrically this is clear, a flat tetrahedron has \textit{faces} that lie in a plane and the normals to these faces are all collinear. Thus it is only the subset of planar configurations of vectors $\vec{A}_l$ that are actually collinear that can have a Minkowki type correspondence with a flat tetrahedron. 

We can say more: the collinear vectors must satisfy closure and so, for some choice of signs we must have $\pm A_1 \pm A_2 \pm A_3 \pm A_4 = 0$. Taking into account the ordering convention $A_1\le A_2 \le A_3 \le A_4$ we can bring the number of cases down to just two
\begin{equation}
  A_2 -A_1=A_4-A_3 \quad \text{or}  \quad A_2+A_1=A_4-A_3.
\end{equation}
 The latter condition leads to a trivial shape space consisting of a single point because $A_{\text{min}}=A_{\text{max}}$. We will call these two conditions the ``flatness" conditions. They are also significant for the Bohr-Sommerfeld quantization. 

Note that, even for collinear configurations of the $\vec{A}_l$ the Minkowski theorem still doesn't hold. The trouble is uniqueness. An infinite number of flat configurations all share the same area vectors. In fact, the differential structure of the shape space breaks down when the flatness conditions are satisfied (for an analogous observation see \cite{Littlejohn:2009}). The qualitative picture is that when flat configurations are present the phase space sphere develops a cusp and looks more like an inverted rain drop. 

 Notice that everything that has been said up to this point is in regards to a tetrahedron that is exactly flat. This is significant because it highlights the singular nature of the Minkowski construction for planar configurations of the $\vec{A}_{l}$. One can construct a tetrahedron with arbitrarily small volume from a set of vectors that is arbitrarily close to planar, there is only trouble when exactly flat tetrahedra are desired. The precise treatment of flat configurations warrants further investigation. 

For the purposes of the present work we summarize the preceding observations: unless the flatness conditions are satisfied, $Q=0$ should not lead to the conclusion that there is a constructible tetrahedron with $V=0$; if a flatness condition is satisfied then only the corresponding pole of the shape space sphere corresponds to flat configurations with $V=0$ and it generally corresponds to a whole class of such tetrahedra\footnote{A simple example: Consider the flat tetrahedron with four equal areas $A_l = \frac{1}{2} a^2 \  (l=1,\dots, 4)$. One example would look like a square made out of two right triangles of side length $a$ and with, say, upward pointing normals and two downward normals. However, this can also be achieved with any rhombus that has side length $b$ and acute angle $\beta$ satisfying $b^2 \sin{\beta} = a^2$. Thus there is a one parameter family of tetrahedra with $A_l = \frac{1}{2} a^2$. }; thus the great circle on which $Q=0$ ($\phi \in \{0,\pi\}$) will be regarded as mathematically useful but largely physically meaningless.

Turning to the maxima of $Q$, take three faces to have fixed areas, say $A_1, A_2$ and $A_3$ but the full vectors $\vec{A}_s, (s=1,2,3)$ not to be given. Writing the triple product of $Q$  as the determinant of a matrix $M=(\vec{A}_1,\vec{A}_2, \vec{A}_3)$ whose columns are the vectors $\vec{A}_1, \vec{A}_2$ and $\vec{A}_3$ and squaring yields,
\begin{equation}
Q^{2} = \frac{4}{81} \det{M}^{T}  \det{M} = \frac{4}{81}
\det{\begin{pmatrix}
A_1^{2} & \vec{A}_1 \cdot \vec{A}_2 & \vec{A}_1 \cdot \vec{A}_3\\
 \vec{A}_2 \cdot \vec{A}_1 &A_2^{2} & \vec{A}_2 \cdot \vec{A}_3\\
 \vec{A}_3 \cdot \vec{A}_1 & \vec{A}_3 \cdot \vec{A}_2&A_3^{2} \\
\end{pmatrix}},
\end{equation}
where $M^{T}$ denotes the transpose of $M$. Taking the unknown dot products $\vec{A}_{s} \cdot \vec{A}_t, (s<t=1,2,3)$ as variables and extremizing one finds the minima already discussed, where $\vec{A}_1$, $\vec{A}_2$ and $\vec{A}_3$ are collinear, and a single global maximum where $\vec{A}_{s} \cdot \vec{A}_t  =0, (s<t=1,2,3)$.
This maximum must satisfy closure and so,
\begin{equation}
\label{area4}
A_4^{2} = A_1^{2}+A_2^{2}+A_3^{2}+2 \vec{A}_1 \cdot \vec{A}_2+2 \vec{A}_1 \cdot \vec{A}_3+2 \vec{A}_2 \cdot \vec{A}_3= A_1^{2}+A_2^{2}+A_3^{2}.
\end{equation}
Then the maximum volume of a tetrahedron with three fixed face areas is the one with three right dihedral angles and the fourth face area given by the equation above. If the same technique is used to maximize the volume over the space where all four face areas are given then the closure condition must be implemented as a constraint. For fixed $A_1$, $A_2$ and $A_3$ it is clear that the constrained maximum will not be larger than the one just found and will be equal to it when the fixed value of $A_4$ is that of \eqref{area4}. Rather than implementing the constraint with a Lagrange multiplier it is easier to extremize equation \eqref{edgeVolArea}, which treats the $A_l, (l=1,\cdots 4)$ on an equal footing. Once again we use the shorthand $P_{0}(A^{2})$ for the polynomial part of the area product $\Delta \bar{\Delta}$,
\begin{equation}
\label{explicitP0}
P_{0}(A^{2}) \equiv [A^{2}-(A_1-A_2)^{2}][A^{2}-(A_3-A_4)^{2}][(A_1+A_2)^{2}-A^2][(A_3+A_4)^{2}-A^2],
\end{equation}
that is, $\Delta \bar{\Delta} = 1/16 \sqrt{P_{0}(A^{2})}$. The expression for the squared volume \eqref{edgeVolArea} simplifies to $Q = 1/(18 \sqrt{x}) \sqrt{P_{0}(x)} \sin{\phi}$. This expression is maximized when $\phi=\pi/2$ and when $\partial Q/\partial A=0$ or,
\begin{equation}
\label{eq:MaxCond}
P_{0}(x) = x P_{0}'(x).
\end{equation}
This condition is another quartic equation, the roots of which are not worth explicitly displaying in general, however choosing the root which maximizes $Q$, say $\bar{x}$, then we have $Q_{\text{max}} = 1/18 \sqrt{P_{0}'(\bar{x})}$, which justifies our claim at \eqref{eq:MaxVol}. This expression will be useful below for finding the largest eigenvalues of $\hat{V}$. These two cases exhaust the most natural constraints on the face areas. If you fix only two of the face areas the volume can grow without bound.

So far we have considered only constraints on the face areas. Another natural constraint is to require that the total surface area of the tetrahedron be constant and look for the largest volume within this class. As one might expect, similar arguments to those presented above lead to the conclusion that the largest volume tetrahedron under this constraint is the one with all face areas equal.  This concludes our general treatment of the extrema of $Q$. 

\subsection{Largest eigenvalues}
\label{sec:largest}

The limiting behavior of the volume spectrum for large and small eigenvalues can now be explored with the assistance of the analytical formula \eqref{eq:CompactAction}. Let us first consider large eigenvalues: the volume function attains a maximum on the sphere and so our strategy will be to expand the action function around this maximum. We have,
\begin{equation}
\label{eq:TaylorAction}
S(q) = S(q_\text{max})+\left.\frac{\partial S}{\partial q}\right|_{q=q_{\text{max}}} (q-q_{\text{max}}) + \cdots.
\end{equation}
From the theory of action-angle variables, the derivative of the action with respect to the Hamiltonian (in this case the volume) is the period $T$ of the system. For the largest eigenvalue, we are in the same situation as if we were finding a ground state, that is, instead of capturing  $2 \pi$ worth of area on the sphere this state only captures an area $\pi $ and so,
\begin{equation}
(q_{\text{max}}-q) = \frac{S(q_\text{max})-S(q)}{T(q_{\text{max}})} = \frac{\pi}{T(q_{\text{max}})}.
\end{equation}
Then the largest eigenvalue is given by,
\begin{equation}
\label{maxEig}
q = (q_{\text{max}}-\frac{ \pi}{T(q_{\text{max}})}).
\end{equation}
At equation \eqref{eq:MaxCond} we found that the maximum classical volume is attained when,
\begin{equation}
P_{0}(x) = x P_{0}^{\prime}(x) 
\end{equation}
or more explicitly when,
\begin{equation}
\frac{1}{x} = \frac{1}{x-(A_1-A_2)^{2}}+ \frac{1}{x-(A_3-A_4)^{2}}+\frac{1}{x-(A_1+A_2)^{2}}+\frac{1}{x-(A_3+A_4)^{2}}.
\end{equation}
The roots of this quartic for generic $A_l$ are complicated functions of the $A_l$, however in the case $A_1=\dots=A_4=A_0$ this equation is easily solved and one finds, $x= \frac{4}{3} A_0^{2}$. The maximum classical volume in this case is $q_\text{max}= (2^{3}/3^{7/2}) A_0^{3}$ and the period is $T(q_\text{max})= 18 g K(0) = 18 (\sqrt{3}/(2A_0^{2})) (\pi/2) = 3^{5/2}\pi/(2A_0^{2})$. The maximum eigenvalue is given by putting these values into \eqref{maxEig},
\begin{equation}
v= q^{1/2} = \frac{2^{3/2}}{3^{7/4}}  A_0^{3/2} \sqrt{1 - \frac{3}{4 A_0}}.
\label{max-equiarea}
\end{equation}
This reproduces the $A_{0}^{3/2}$ scaling that has been found in previous works and refines it to the next order. This scaling is plotted as the uppermost line in Figure \ref{EqjPlot}. Further corrections could be developed by retaining more terms in \eqref{eq:TaylorAction}. As discussed in Section \ref{subsec:ClassicalCases}, this is the appropriate scaling to consider under the constraint of fixed total surface area. Consequently, this scaling may be interesting to investigate in more detail within the $U(N)$ framework \cite{Girelli:2005}. 

\begin{figure}[ht]
\begin{center}
   \includegraphics[width=3.2in]{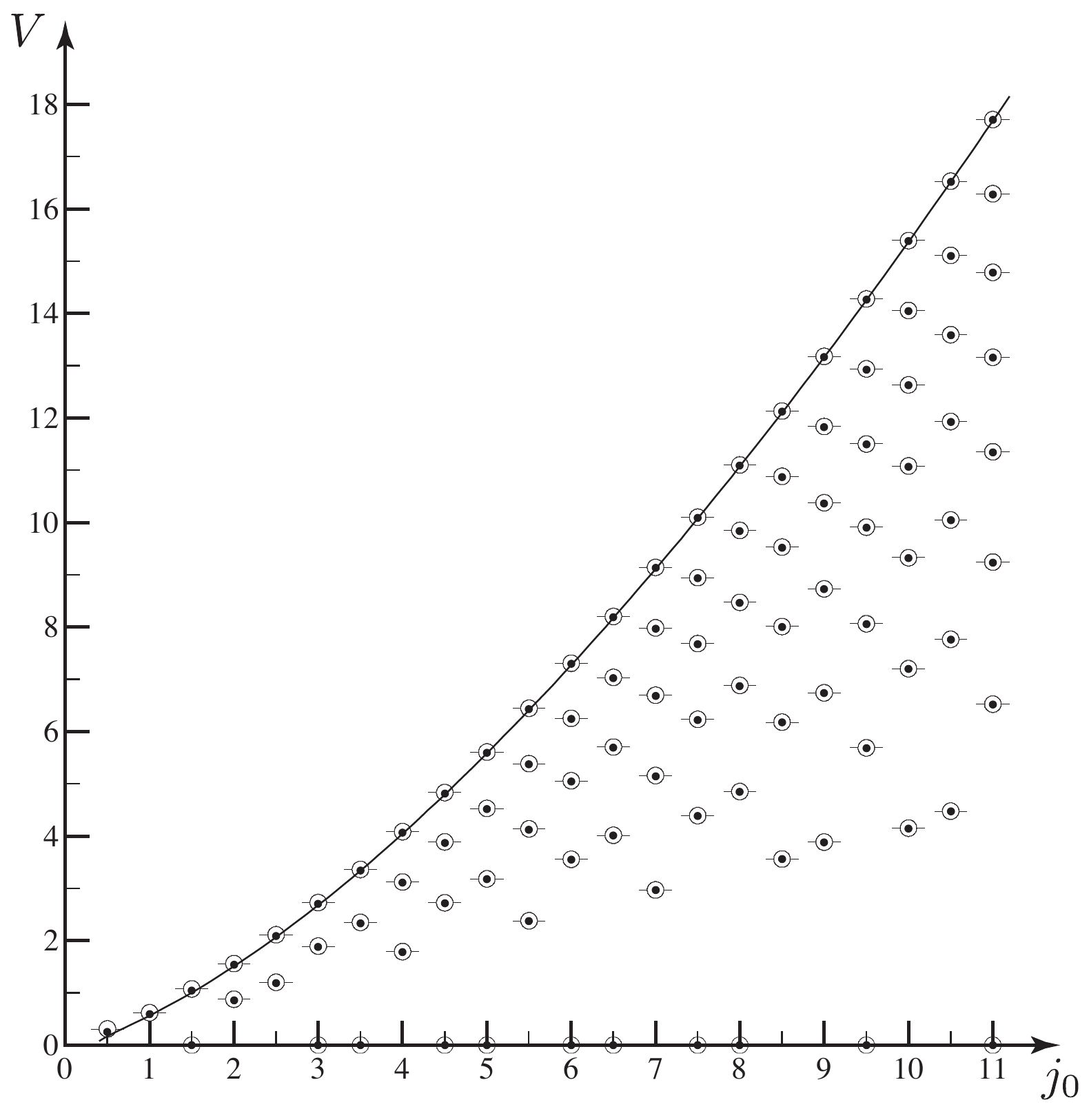}
      \caption[Scaling of the largest volume eigenvalue]{This figure compares the semiclassical scaling of the largest volume eigenvalue (dark line) to the Bohr-Sommerfeld (dots) and loop gravity (circles) spectra. The first three spins are given by $j_1=j_2=j_3=j$ and and the largest spin $j_4$ is given by the closest integer or half integer to $\sqrt{3}j$, that respects the Clebsch-Gordan conditions.}
      \label{NonEqjPlot}
  \end{center}
\end{figure}

Occasionally the equal area tetrahedron has been supposed to be the one whose volume grows most rapidly as the areas are increased. However, we have seen above that this is not always the case. The tetrahedron with maximum volume depends on the space under consideration. For the space with three face areas fixed, corner tetrahedra with, for example, $A_1=A_2=A_3=A_0$ and $A_4=\sqrt{3} A_0$ are maximizing. Indeed, solving \eqref{maxEig} for these corner tetrahedra we find a larger scaling coefficient,
 \begin{equation}
 v = q^{1/2} = \frac{\sqrt{2}}{3} A_{0}^{3/2} \sqrt{1- \frac{\sqrt{3}}{2 A_0}}.
 \label{max-trirect}
 \end{equation}
  In Figure \ref{NonEqjPlot} the Bohr-Sommerfeld spectrum is compared to the numerically calculated exact spectrum and the scaling derived here for the corner tetrahedra. 

\subsection{Smallest eigenvalues: non-flat configurations}

The small eigenvalue case has subtleties associated with it. As is clear from the Taylor expansion, 
\begin{equation}
\label{smallTaylor}
I(q) = I(0)-\left.\frac{\partial I}{\partial q}\right|_{q=0} q + \cdots,
\end{equation}
the smallest eigenvalues are associated with the longest period at zero volume,  $q_\text{min} = (I(0)-I(q))/T(0)$. This would be the end of the story except for the fact that there are a number of shape spaces for which the period at zero volume can become infinitely long and the Taylor expansion above is invalidated. For this reason, we have to treat different shape spaces individually. We will begin by treating the case where the period is finite, as it is simpler.

One subtlety of the small volume cases is immediate: the smallest eigenvalue depends on the dimension of the intertwiner space, $d \equiv \dim{\IS[4]}$. As already noted, the volume is odd under parity and so there is only a zero eigenvalue when $d$ is odd and this is the only volume state invariant under parity. We are interested in the first non-zero eigenvalue and thus when $d$ is odd the spacing $I(0)-I(q)$ is the spacing between two quantized orbits, $h$, or in our units $2\pi$. For the finite period case that we are considering the Taylor expansion \eqref{smallTaylor} is valid and we have, $T(0) q_{\text{min}} = 2\pi$. On the other hand if the phase space is even dimensional, neighboring quantized orbits evenly straddle the zero volume contour and we have $I(0)-I(q)=\pi$, so that $T(0) q_{\text{min}} = \pi$.

We found the period above, in general it is $T = 18 g K(m)$, where $g$ and $m$ are given in equations \eqref{ellipticMod} and \eqref{AsqLam}. For $q=0$ the quartic $P(x,Q=0)$ factorizes and the roots are $\bar{r}_{1}=(A_1-A_2)^{2}, \bar{r}_{2}=(A_3-A_4)^{2}, \bar{r}_{3} = (A_1+A_2)^{2}, \bar{r}_{4}= (A_{3}+A_4)^{2}$, for example. Of course, depending on the choice of $A_1, \dots, A_4$ other orderings are possible (recall that the $r$ are defined such that $r_1<r_2<r_3<r_4$). If, as suggested after equation \eqref{AsqLam}, $m$ is always chosen such that $m<1$ these other orderings lead to the same result: $g= 1/(2 \sqrt{A_1A_2A_3A_4})$  and 
\begin{align}
\begin{aligned}
\label{matQzero}
\qquad m &= \frac{(A_1+A_2+A_3-A_4)(A_1+A_2-A_3+A_4)}{(2A_1)(2A_2)}\\
&\qquad \times \frac{(A_1-A_2+A_3+A_4)(-A_1+A_2+A_3+A_4)}{(2A_3)(2A_4)}.
\end{aligned}
\end{align}
The complete elliptic integral of the first kind has the power series expansion,
\begin{equation}
\label{EllInt1}
K(m) = \frac{\pi}{2} \sum_{0}^{\infty} \left[\frac{(2n)!}{2^{2n} n!^{2}}\right]^{2} m^{n},
\end{equation}
valid for $m<1$. If $m$ is small enough such that higher order terms can reasonably be neglected then, for $d$ odd,
\begin{equation}
q_{\text{min}} \approx \frac{2 \pi}{9\pi/2\sqrt{A_1A_2A_3A_4}} = \frac{4}{9} \sqrt{A_1 A_2A_3A_4},
\end{equation}
and this can be improved as much as desired by including more terms from \eqref{EllInt1}. Expressed in terms of the volume of the tetrahedron we have,
\begin{equation}
\label{eq:minNoFlats}
v_{\text{min}} \approx
(A_1A_2A_3A_4)^{1/4} \begin{cases}& 2/3 \quad \text{if $d$ is odd}\\
& \sqrt{2}/3 \quad \text{if $d$ is even},
\end{cases}  \qquad \text{for $m \ll 1$}.
\end{equation}

\subsection{Smallest eigenvalues: flat configurations}

The exact volume eigenvalues derived by Brunnemann and Thiemann \cite{Brunnemann:2006} are for special cases of the $A_1, \cdots, A_4$, such as $A_1=A_2=1$ and $A_3=A_4=j+1/2$, where only two of the four vectors grow as you increase $j$. For cases like these we find the same qualitative scaling from the formula \eqref{eq:minNoFlats}, $v \sim j^{1/2}$. However, notice that this special choice of the $A_r$ leads to $m=1$ and invalidates the expansion of the elliptic integral $K(m)$, in particular $K$ logarithmically diverges (hence also the period) and a different approach to estimating the eigenvalues is necessary. Brunnemann and Thiemann were led to consider these special cases by their numerics, they found that these were the phase spaces which lead to the smallest overall values for the volume spectrum. Below we describe what is special about the geometry of these cases and develop an alternative technique for estimating the spectra of these spaces. 

The longest periods are achieved when $m=1$ and the elliptic function theory limits to elementary functions, for example, the standard roots of the Jacobi form of the elliptic functions, $\pm 1$ and $\pm1/\sqrt{m}$, degenerate. We worked out the elliptic modulus in the zero volume limit above, see \eqref{matQzero}. Setting this expression equal to 1 we find four roots, which can be regarded as expressing any one of the $A_r$ in terms of the other three, these are:
 $A_{1} - A_{2} = A_{4}-A_{3}$, $A_{1} +A_{2} = A_{3}+A_{4}$, $A_{1} + A_{2} = -A_{3}-A_{4}$ and $A_{2} - A_{1} = A_{4}-A_{3}$, only the last of which is physical due to our ordering convention on the $A_l$. This final condition is the first of the flatness conditions introduced in Section \ref{subsec:ClassicalCases}. (The other flatness condition arises when you consider the $m=0$ limit.) Remarkably, the elliptic function theory limits to elementary functions precisely when the shape space contains flat configurations. This provides a geometrical interpretation to the special cases that Brunneman and Thiemann investigated\footnote{The flatness condition explains most of their cases. A few cases are special and not due to the flatness condition but rather because the dimension of the space of intertwiners is two.} \cite{Brunnemann:2006}: they are special because of contributions from flat tetrahedra. 
 
\begin{figure}[h]
\begin{center}
   \includegraphics[height=3.5in]{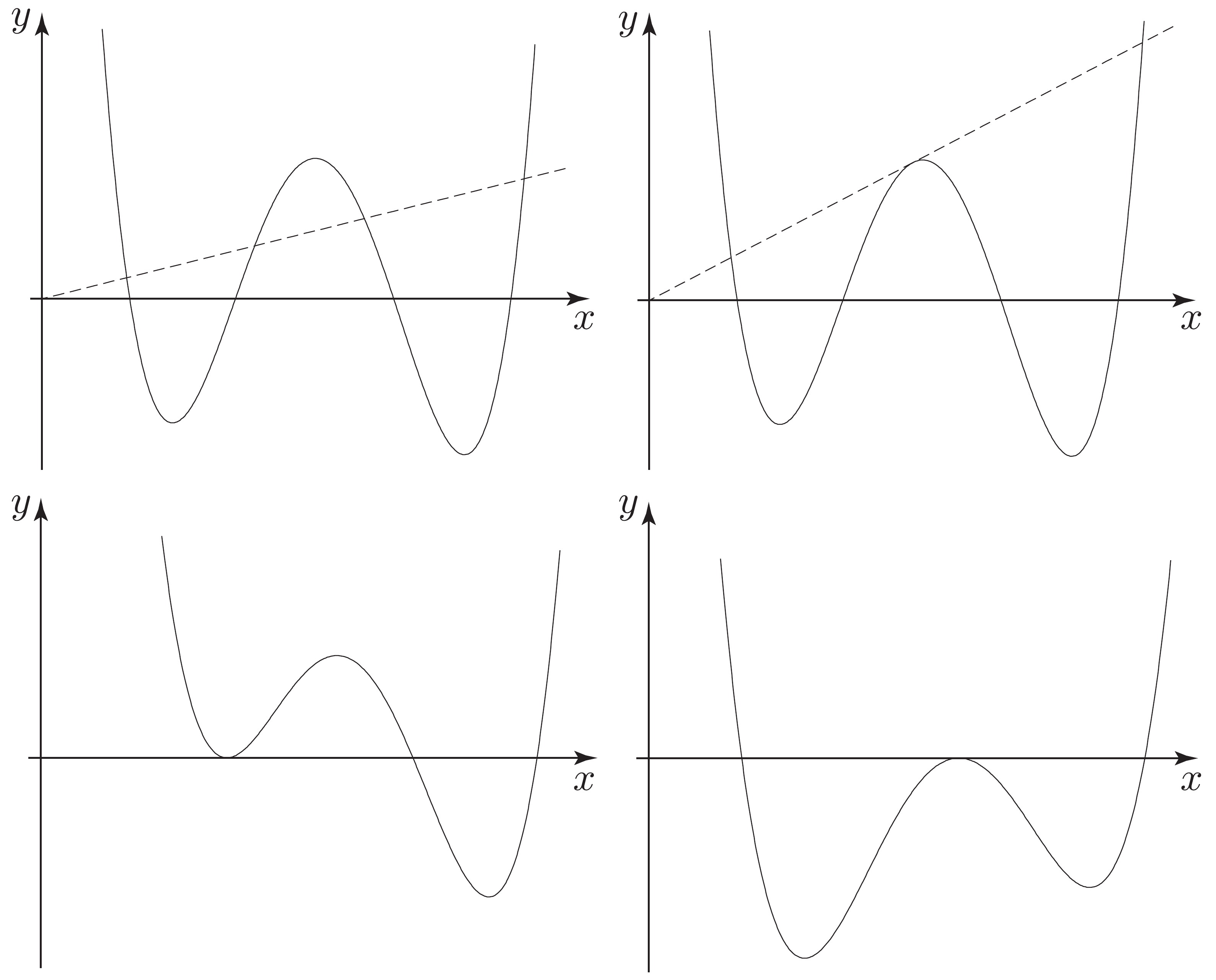}
      \caption[Characterization of shape space by a quartic]{This figure summarizes our characterization of the shape space using the quartics $P_{0}(x)$ and $P(x)$, defined by equations \eqref{explicitP0} and \eqref{eq:PDef}. The straight lines are given by $y=324 Q^{2} x$ and the quartic curves by $y=P(x,0)\equiv P_{0}(x)$. The upper left panel shows the generic case in which the fixed volume determines four distinct roots. The upper right panel shows the coalescence of the middle roots at maximum volume and $m=0$. The lower two panels show the coalescence of other pairs of roots at zero volume and $m=1$ (these are two distinct cases).}
      \label{QuartGrid}
  \end{center}
\end{figure}

 This flatness condition also has a nice interpretation in terms of the polynomial $P_{0}(x)$; it indicates that two of the polynomials roots are coalescing. Put this together with our observation that the maximum volume is achieved when $r_{2}=r_{3}$ and, consequently, when $m=0$ and we find that $P_{0}(x)$ is quite useful for characterizing the shape space of a tetrahedron with fixed face areas. These findings are summarized in Figure \ref{QuartGrid}.

In the case where $m=1$ we can no longer use the Taylor expansion and elliptic function period to find the eigenvalues, instead we have to find the small volume behavior of the action function and try to invert it. The action depends on the volume in two ways, an explicit dependence through the prefactor $Q$ in \eqref{finalAction}, and an implicit dependence through the roots of the quartic $P(x,Q)$. Thus the first step in finding the small volume behavior is to expand the roots as a power series in $Q$. This would be quite laborious if we had to go through the solutions to the quartic equation, happily a simple alternative exists because we know the roots $\bar{r}_i$ at $Q=0$. We simply plug $\bar{r}_i+\rho$ into $P(x,Q)$ and require that $\rho$ is such that the equation is satisfied at lowest order in $Q$, this process can be iterated to find $r_{i}$ to the desired order in $Q$. The result of these calculations to fourth order in $Q$ are summarized in Appendix B.

Using the series expansions of the roots we can Taylor expand the action as a power series in $Q$. This is slightly delicate because the complete elliptic integrals diverge logarithmically at $m=1$, however, taking care to expand the logarithms to the proper order we find simple results. First we consider the case in which the $A_{l}$ are all equal with $A_l \equiv A_0$, so that $\bar{r}_1=\bar{r}_{2}\equiv 0$ and $\bar{r}_{3}=\bar{r}_{4}\equiv \tilde{r} = (2 A_0)^{2}$. The action simplifies and the expansion yields,
\begin{align}
I &= 18g Q\left(\left[-1-2 \frac{r_{4}  }{(r_{4}-\tilde{r})}\right]K(m)-2 \frac{\tilde{r}(r_{4}-r_{3})}{(r_{4}-\tilde{r})(r_{3}-\tilde{r})}\Pi(\alpha_{4}^{2}, m)\right)\\
&\approx \sqrt{\tilde{r}} \pi + \frac{18 Q}{\tilde{r}} \ln{\left(\frac{9 Q}{e \tilde{r}^{3/2}}\right)^{3}} +O(Q^{2}(\ln{Q}))\\
\label{twoEqRoots}
&= I(0)+(6 e \tilde{r}^{1/2})\left(\frac{9 Q}{e \tilde{r}^{3/2}}\right) \ln{\left(\frac{9 Q}{e \tilde{r}^{3/2}}\right)},
\end{align}
where in the last equality we have recognized $\sqrt{\tilde{r}} \pi= 2 \pi A$ as half of the symplectic area of the sphere and hence the action at zero volume squared, $I(0)$. 

\begin{figure}[h]
\begin{center}
   \includegraphics[height=3in]{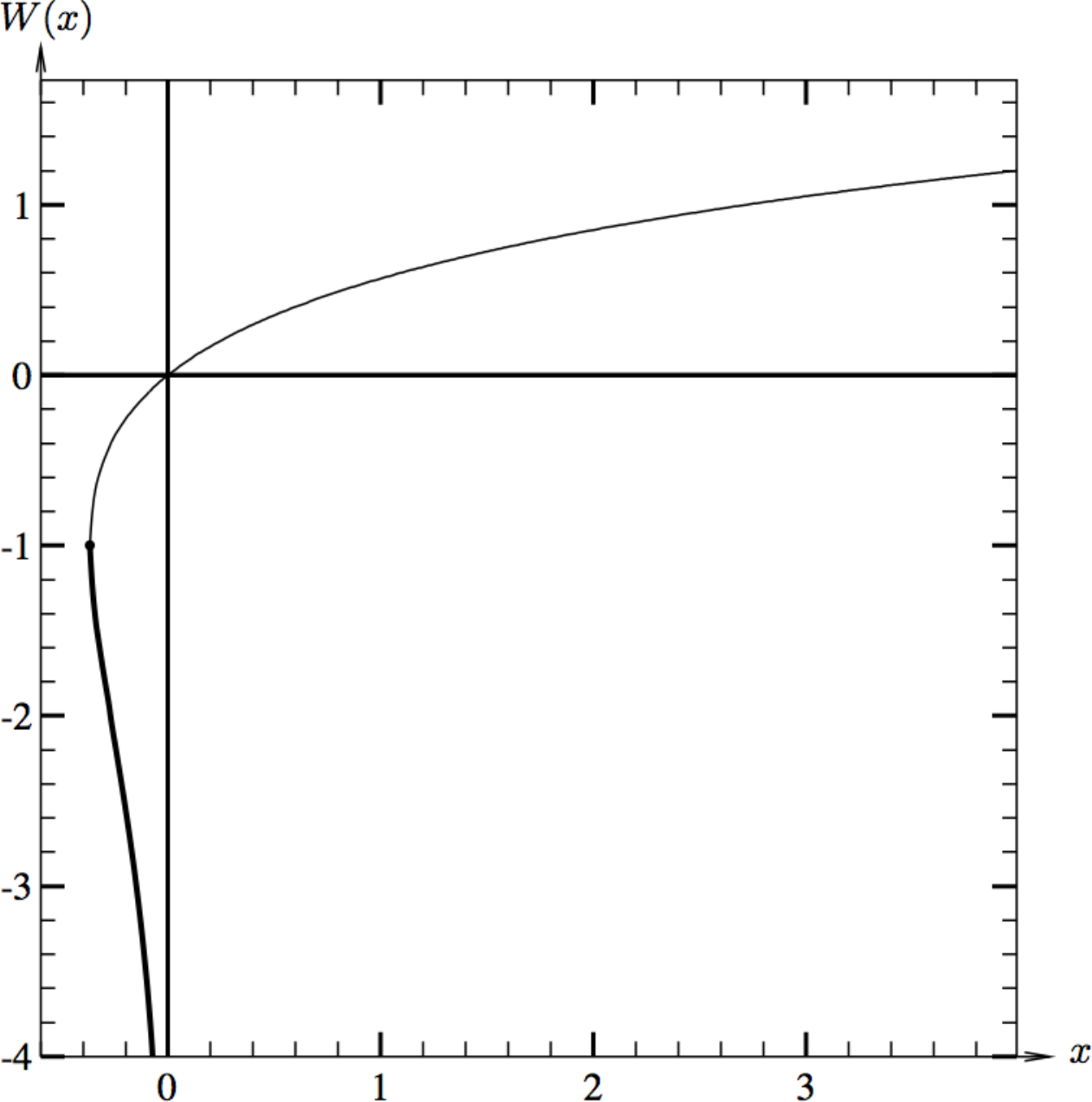}
      \caption{The real values of the Lambert W function}
        \label{fig:LambertW}
  \end{center}
\end{figure}

At lowest order then, the relation between $I$ and $Q$ can be inverted using Lambert's $W$ function. Because this inverse function is uncommon, we briefly review the properties used in this work. The Lambert $W$ is defined as the function which inverts the relationship,
\begin{equation}
x = W e^{W},
\end{equation}
yielding $W(x)$. The function $W(x)$ is plotted in Figure \ref{fig:LambertW}.
Note that the function is multivalued in the interval $[-1/e, 0]$, the upper branch (lighter shade in Figure \ref{fig:LambertW}) is conventionally taken to be the principle branch $W_{0}(x)$. However, for the expansion that we are interested in, we need the lower branch $W_{-1}(x)$ (darker shade in Figure \ref{fig:LambertW}). As $x$ approaches zero from below $x \rightarrow 0_{-}$ the lower branch $W_{-1}(x)$ can be developed in the following series,
\begin{equation}
W_{-1}(x) \approx  - \ln{(-\frac{1}{x})}-\ln{(\ln{(-\frac{1}{x})})}- \frac{\ln{(\ln{(-\frac{1}{x})})}}{\ln{(-\frac{1}{x})}} + \cdots,
\end{equation}
this is the series that we will need to complete our derivation of a lower bound on the volume spectrum. 

\begin{figure}[ht]
\begin{center}
   \includegraphics[width=3.2in]{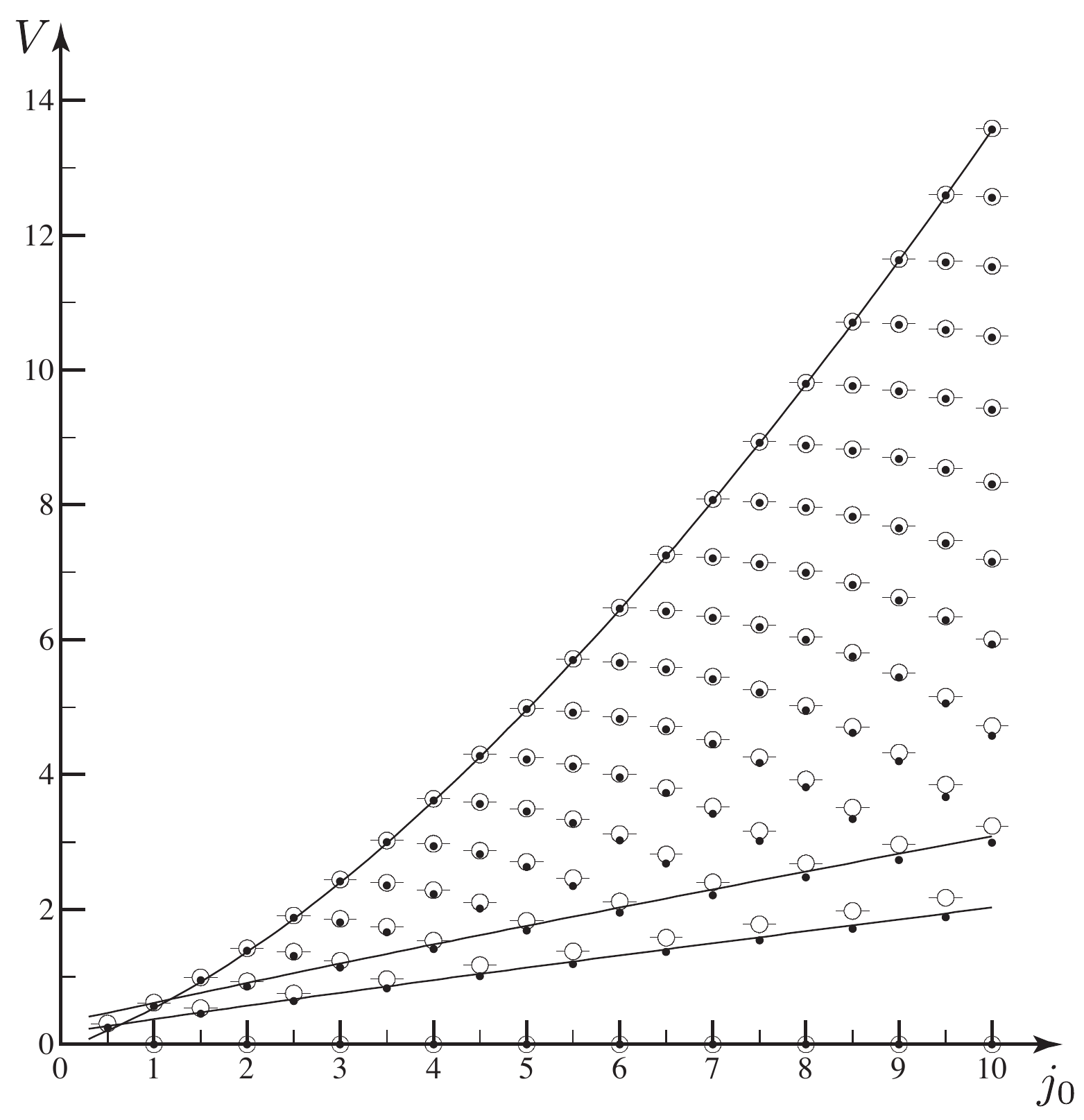}
      \caption[Scaling of the largest and smallest volume eigenvalues]{This figure compares the semiclassical scaling of the largest and smallest volume eigenvalues (dark lines) to the Bohr-Sommerfeld (dots) and loop gravity (circles) spectra. The four spins are equal,  $j_1=j_2=j_3=j_4=j_0$ , and the corresponding phase space contains flat configurations. This explains the poorer agreement of the Bohr-Sommerfeld spectra at small eigenvalues. }
      \label{EqjPlot}
  \end{center}
\end{figure}

Using these observations about the Lambert W function we have from \eqref{twoEqRoots},
\begin{equation}
Q \approx \frac{\tilde{r} (I-I_{0})}{54 W_{-1}((I-I_{0})/(6 e \tilde{r}^{1/2}))}.
\end{equation}
And expanding this solution for small $(I-I_{0})/(6 e \tilde{r}^{1/2})$ yields,
\begin{equation}
\label{logSeries}
V \approx \frac{2\sqrt{\pi}}{3 \sqrt{3}} A \left(\frac{1}{\sqrt{\ln{\left( \frac{6 e A}{\pi} \right)}+\ln{\left(\ln{\left( \frac{6 e A}{\pi} \right)}\right)} +\cdots } } \right),
\end{equation}
in the case that the phase space is odd dimensional, and a very similar result in the even dimensional case. Both cases are plotted in Figure \ref{EqjPlot} the odd dimensional case being the middle curve and the even dimensional case the lowest curve. Note that the smallest Bohr-Sommerfeld eigenvalues in this plot are in much poorer agreement than for the case of Figure \ref{EqjPlot}. This is due to the fact, discussed in section \ref{subsec:ClassicalCases}, that the space of shapes is no longer a differentiable manifold when we consider equal spins $j_l$, i.e. equal areas $A_l$.  

\section{Conclusion}
\label{sec:VolConcs}

At the Planck scale, a quantum behavior of the geometry of space is expected. Loop gravity provides a specific realization of this expectation: it predicts a granularity of space with each grain having a quantum behavior. In this paper we have presented a new independent road to the granularity of space and the computation of the spectrum of the volume. The derivation is based on semiclassical arguments applied to the simplest model for a grain of space, a Euclidean tetrahedron, and is closely related to Regge's discretization of gravity and to more recent ideas about general relativity and quantum geometry \cite{Freidel:2010a}. The spectrum has been computed by applying Bohr-Sommerfeld quantization to the volume of a tetrahedron seen as an observable on the phase space of shapes. We briefly summarize our results:
\begin{itemize}
\item[i.] \emph{Spectrum}. We studied volume orbits on the phase space of a tetrahedron. The main result is a closed formula for the action integral $I(q)$ in terms of elliptic functions, eq. (\ref{eq:CompactAction}). This formula is then used to compute the Bohr-Sommerfeld levels of the volume. In Figures \ref{fig:Comparison}, \ref{NonEqjPlot} and \ref{EqjPlot} and in the tables in Appendix \ref{app:tables} we compare the volume levels to the spectrum computed in loop gravity and find a remarkable quantitative agreement.
\item[ii.] \emph{Degeneracy}. Non-vanishing eigenvalues of the volume are twice generate. From the semiclassical perspective, this is understood as a consequence of the fact that there are two closed orbits with the same volume. Tetrahedra on the two orbits are related by parity. Quantized orbits with zero volume appear only when the symplectic area of phase space is an odd multiple of the elementary, phase-space Planck cell. In this case parity sends the orbit into itself and the associated eigenvalue is non-degenerate.\\
In section  \ref{subsec:ReggeSymmetry} we also identified a new degeneracy that connects multiplets in different intertwiner spaces. At the semiclassical level this symmetry originates from scissor-congruent auxiliary tetrahedra, and at the quantum level from the Regge symmetry of the $\{6j\}$ symbol. We conjecture that the associated geometric tetrahedra are scissor-congruent as well.
\item[iii.] \emph{Maximum volume}. At the classical level, fixing the four areas $A_1,\ldots,A_4$, there is a maximum volume the tetrahedron can attain. In section \ref{sec:largest} we computed the largest eigenvalue of the volume  by expanding the action integral $I(q)$ around the maximum classical volume. The largest eigenvalue is smaller than the maximum classically-allowed volume and the difference between the two is given by $\pi$ divided by the period $T$ of the volume orbit. Moreover the large eigenvalues are equispaced with a separation $2\pi/T$. This phenomenon can be understood as a manifestation of Bohr's correspondence principle.\\
 If instead of fixing the area of each of the four faces we fix only the area of three faces, then the tetrahedron of maximum volume is tri-rectangular and the maximum eigenvalue is given in equation \eqref{max-trirect}. We also considered fixing only the total surface area. In this case the maximum volume is attained by a tetrahedron with faces having all the same area and the maximum eigenvalue is given in equation \eqref{max-equiarea}. This formula reproduces the scaling $v_{\text{max}}\sim j^{3/2}$ found by Brunnemann and Thiemann, and determines corrections to it. 
\item[iv.] \emph{Minimum volume}. We studied the volume gap for a quantum tetrahedron by pushing the Bohr-Sommerfeld approximation into the deep quantum regime. The elliptic parameter $m\ll1$ that controls the accuracy of the approximation is given by equation \eqref{eq:minNoFlats}. For phase spaces containing no flat configurations, we find a volume gap $v_{\text{min}}=c\,(A_1 A_2 A_3 A_4)^{1/4}$, with $c$ equal to, $2/3$ for an odd  and $\sqrt{2}/3$ for an even, number of levels.\\
 We have also studied phase spaces containing flat configurations. An example is the equi-area case. This situation is more delicate as there are orbits of infinite period corresponding to singular points in phase space. Physically this is also the most interesting situation as it leads to the smallest attainable volume when the areas are all equal to the smallest area $A=1/2$. All our results support the existence of a volume gap for the 4-valent case given in the equi-area case by the unusual Log series of eq. \eqref{logSeries}, see Fig. \ref{EqjPlot}.
\end{itemize}

The remarkable quantitative agreement of the spectrum calculated here and the spectrum of the volume in loop gravity lends further credibility to the structure of this theory. The semiclassical methods of this paper provide a new understanding of many aspects of the rich structure of the volume spectrum in loop gravity and the explicit formulas open new avenues for analytical exploration. This is important because a deep understanding of the spectra of geometrical operators provides fertile ground for developing phenomenological tests of loop gravity \cite{Major:2010}.

\section{Acknowledgements}
We would like to thank R. Littlejohn, C. Rovelli for discussion and inspiration. HMH thanks University of California, Berkeley for fellowship support and the Perimeter Institute for hospitality and support during the completion of this work. Research at Perimeter Institute for Theoretical Physics is supported in part by the Government of Canada through NSERC and by the Province of Ontario through MRI. 

\appendix
\section{Derivation of volume matrix elements}

The simplest derivation of the matrix elements of the volume operator that we know of is also the oldest and is due to L\'evy-Leblond and L\'evy-Nahas, \cite{LevyLeblond:1965}. The presentation in this Appendix closely parallels their argument. 

Note first that
\begin{equation}
[\vec{J}_1 \cdot \vec{J}_2, \vec{J}_1 \cdot \vec{J}_3] = [J_{1i}, J_{1j}] J_{2}^{i} J_{3}^{j} = i \epsilon_{ijk} J_{1}^{k}J_{2}^{i} J_{3}^{j}  = i \vec{J}_{1} \cdot (\vec{J}_2 \times \vec{J}_{3}),
\end{equation}
which is also $[\vec{J}_1 \cdot \vec{J}_2, \vec{J}_1 \cdot \vec{J}_3]=\frac{1}{2}[(\vec{J}_1+\vec{J}_2)^2, \vec{J}_1 \cdot \vec{J}_3]$. This allows one to express the matrix elements of $\hat{Q}$ by
\begin{equation}
\label{eq:QasJ13}
\begin{aligned}
\langle k| \hat{Q} | k' \rangle \equiv Q_{k}^{\phantom{k}k'} &= \langle k| \vec{J}_{1} \cdot (\vec{J}_2 \times \vec{J}_{3}) | k' \rangle =- \frac{i}{2} \langle k| \frac{1}{2}[(\vec{J}_1+\vec{J}_2)^2, \vec{J}_1 \cdot \vec{J}_3] | k' \rangle\\
& = -\frac{i}{2} (k(k+1)-k'(k'+1)) \langle k| \vec{J}_1 \cdot \vec{J}_3 |k' \rangle,
\end{aligned}
\end{equation}
where the last equality follows from our definition of the $|k \rangle$ basis (Eq. \eqref{eq:|k>} and above).  Already it is clear that the diagonal matrix elements vanish. Then, the problem has been reduced to evaluating the matrix element $\langle k | \vec{J}_1 \cdot \vec{J}_3 | k' \rangle$ for $k \neq k'$. Evaluating this matrix element amounts to two applications of the Wigner-Eckart theorem, which are performed here graphically. The graphical notation is quite efficient as long as you do not track phases. The overall phase is not needed anywhere in this work and so we proceed without tracking phases, our end result is in agreement with \cite{LevyLeblond:1965}.

Inserting two resolutions of the identity yields, 
\begin{equation}
\langle k | \vec{J}_1 \cdot \vec{J}_3 | k' \rangle = a_0(k,k') \times  \raisebox{-.62in}{ \includegraphics[height=1.3in]{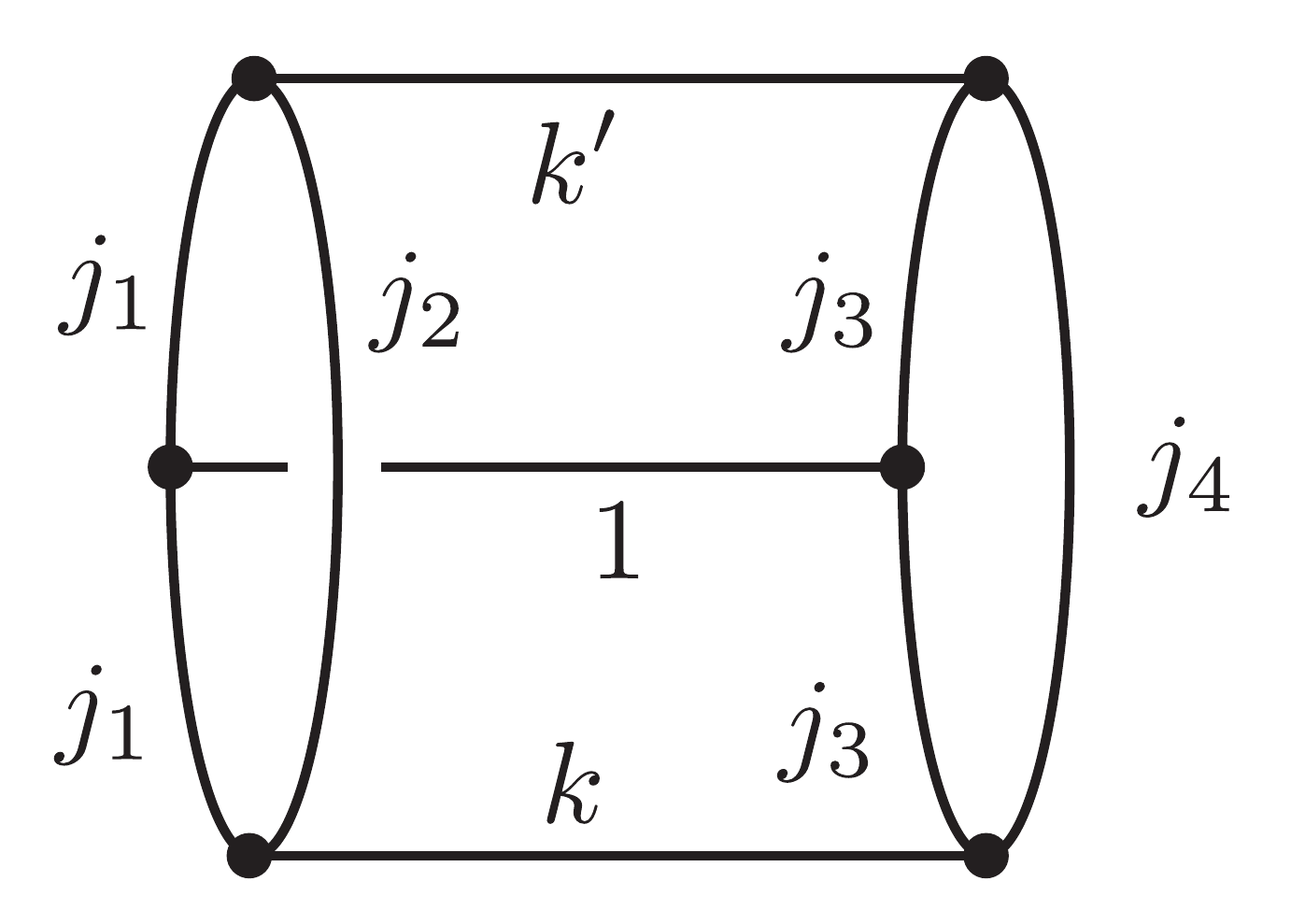} }
\end{equation}
where the constant $a_0$ is given by,
\begin{equation}
a_0(k,k') = \sqrt{2 k +1} \sqrt{2k'+1}   \sqrt{j_1(j_1+1)(2j_1+1)}  \sqrt{j_3(j_3+1)(2j_3+1)}.
\end{equation}
Graphs separable on three lines can be simplified, \cite{Yutsis:1962}, and in this case the reduction yields,
\begin{equation}
\label{J13Mat}
\langle k | \vec{J}_1 \cdot \vec{J}_3 | k' \rangle = a_0(k,k') \times  \raisebox{-.62in}{ \includegraphics[height=1.3in]{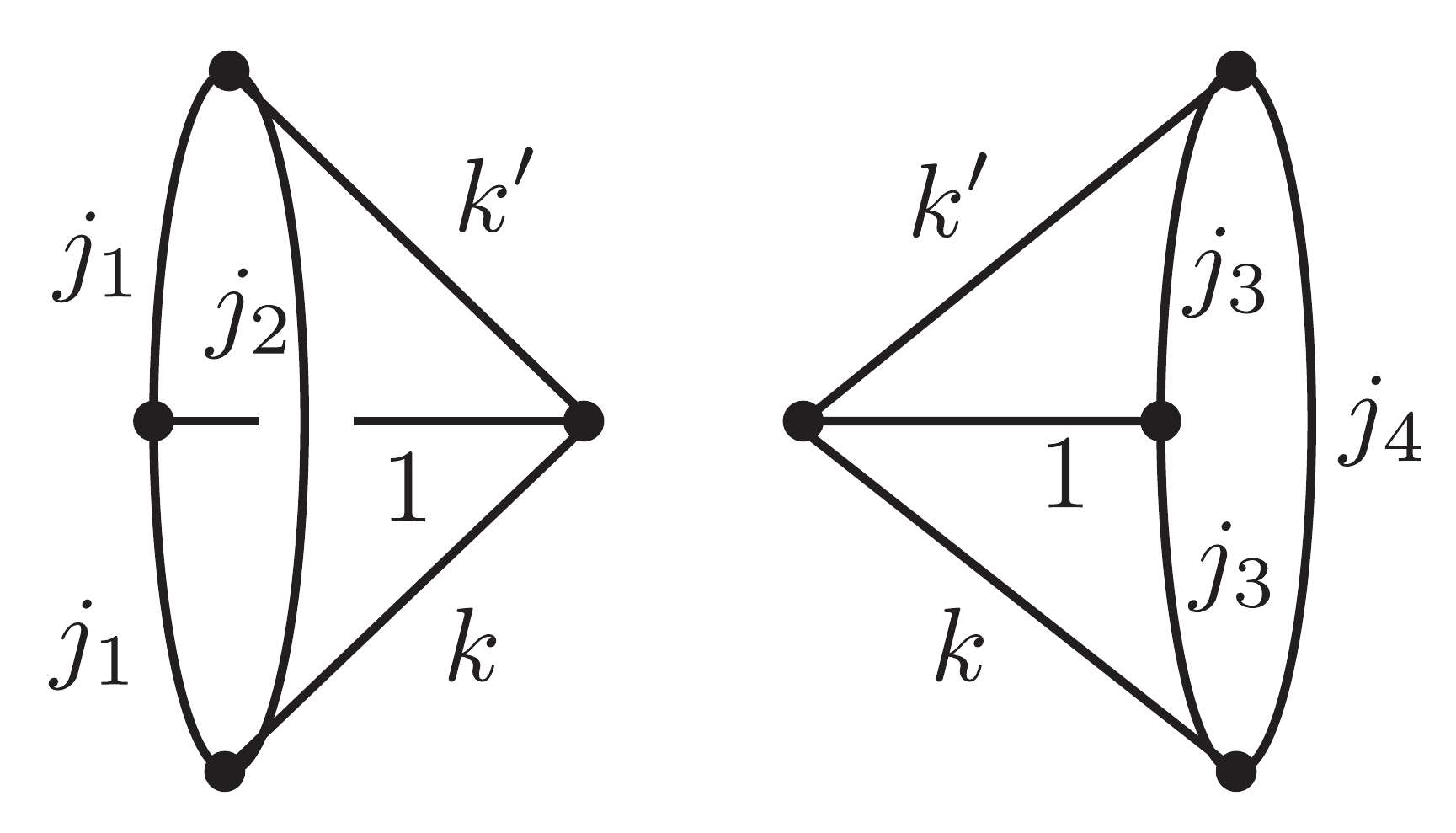} }
\end{equation}
Each of the simplified graphs is a $6j$-symbol and it immediately follows that the matrix elements of the volume operator satisfy selection rules: Each of the nodes at the center of the diagram imposes the condition that the matrix element vanishes unless $k$ and $k'$ differ by one. This implies that the matrix is of the form
\begin{equation}
Q_k^{\phantom{k}k'} = 
\left(\begin{array}{cccc}
0 & i a_1 & 0 &\cdots \\
-i a_1 & 0& i a_2	&\ddots \\
0 & -i a_2 & 0 &\ddots \\
\vdots & \ddots & \ddots& \ddots
\end{array}\right),
\end{equation}
here the real $a_{k}$ are defined by,
\begin{equation}
a_{k-1} = i Q_{k}^{\phantom{k} k-1}  \qquad (k=2,\dots, d+1),
\end{equation}
where, once again, $d$ is the dimension of the intertwiner space. Combining the results from \eqref{eq:QasJ13} and \eqref{J13Mat} it follows that the $a_k$ are given by,
\begin{equation}
\begin{aligned}
a_{k-1} & =\left| \frac{1}{2} (k(k+1)-(k-1)k) a_0(k,k-1) 
\begin{Bmatrix}
k-1 & 1 & k \\
j_1 & j_2 & j_1
\end{Bmatrix}
\begin{Bmatrix}
1 & k-1 & k \\
j_4 & j_3 & j_3
\end{Bmatrix}
\right|
\\[2em]
& = \frac{1}{4} \frac{\sqrt{(j_1+j_2+k+1)(j_1+j_2-k+1)(j_1-j_2+k)(j_2-j_1+k)}}{\sqrt{2k+1}} \\[1em]
&\phantom{blahblah} \times \frac{\sqrt{(j_3+j_4+k+1)(j_3+j_4-k+1)(j_3-j_4+k)(j_4-j_3+k)}}{\sqrt{2k-1}}.
\end{aligned}
\end{equation}

\section{Root Power Series}
The roots $r_i$ $(i=1,\dots, 4)$ are defined as the solutions to the quartic (in $A^2$) polynomial equation,
\begin{equation}
 P(A^{2}, Q^{2}) = (4 \Delta)^{2} (4 \bar{\Delta})^{2}- (2A)^{2} (9Q)^{2} \equiv P_{0}(A^{2})- (2A)^{2} (9Q)^{2}=0,
\end{equation}
where,
\begin{equation}
P_{0}(A^{2}) \equiv [A^{2}-(A_1-A_2)^{2}][A^{2}-(A_3-A_4)^{2}][(A_1+A_2)^{2}-A^2][(A_3+A_4)^{2}-A^2].
\end{equation}
The barred roots $\bar{r}_i$ $(i=1,\dots,4)$ are defined as solutions to $P_{0}(A^2)=0$. 

For the case where the $\bar{r}_{i}$ are distinct we have,
\begin{equation}
r_{i} = \bar{r}_{i} + \lambda_{i} Q^{2} +\beta_{i} \lambda_{i}^{2} Q^{4}+\cdots,
\end{equation}
where 
\begin{equation}
\lambda_{i} \equiv \frac{ 18^{2}  \bar{r}_{i}}{\prod_{j \ne i}(\bar{r}_{i}-\bar{r}_{j})} \qquad \text{and} \qquad \beta_{i} \equiv \left( \frac{1}{\bar{r}_{i}} - \sum_{j\ne i} \frac{1}{\bar{r}_{i}-\bar{r}_{j}} \right), \qquad (i,j=1,\dots,4).
\end{equation}
When $\bar{r}_{1}$ and $\bar{r}_{2}$ coincide, say at $\bar{r}$, the above series are singular and are instead replaced by,
\begin{equation}
r_{i} = \bar{r} \pm \mu Q+\nu Q^{2} \pm \rho Q^{3} + \sigma Q^{4} \pm \cdots \qquad (i=1,2),
\end{equation}
the lower signs for $r_{1}$ and the upper for $r_{2}$ and with,
\begin{align}
\mu &= \frac{18 \sqrt{\bar{r}}}{\sqrt{(\bar{r}_{3}-\bar{r})(\bar{r}_{4}-\bar{r})}}, \qquad \nu = \frac{(18)^{2} (\bar{r}_{4} \bar{r}_{3}-\bar{r}^{2})}{2 (\bar{r}_{3}-\bar{r})^{2}(\bar{r}_{4}-\bar{r})^{2}},\\
\rho &= \frac{(18)^{3} (\bar{r}_{3}^{2} \bar{r}_{4}^{2} +4 \bar{r} \bar{r}_{3} \bar{r}_{4}(\bar{r}_{3}+\bar{r}_{4})-14 \bar{r}^{2} \bar{r}_{3} \bar{r}_{4}+5\bar{r}^{4})}{8 \sqrt{\bar{r}}  (\bar{r}_{3}-\bar{r})^{7/2}(\bar{r}_{4}-\bar{r})^{7/2}}, \\
 \sigma &= \frac{(18)^{4} (\bar{r}_{3}^{2} \bar{r}_{4}^{2}(\bar{r}_{3}+\bar{r}_{4}) +\bar{r} \bar{r}_{3} \bar{r}_{4} (\bar{r}_{3}-\bar{r}_{4})^{2} - 5\bar{r}^{2} \bar{r}_{3} \bar{r}_{4}(\bar{r}_{3}+\bar{r}_{4})+10 \bar{r}^{3} \bar{r}_{3} \bar{r}_{4}-2\bar{r}^{5})}{2 (\bar{r}_{3}-\bar{r})^{5}(\bar{r}_{4}-\bar{r})^{5}}.
\end{align}
If $\bar{r}_{3}$ and $\bar{r}_{4}$ coincide, say at $\tilde{r}$, the above formula holds with the replacements $\bar{r}_{3} \rightarrow \bar{r}_{1}$, $\bar{r}_{4} \rightarrow \bar{r}_{2}$ and $\bar{r} \rightarrow \tilde{r}$. There is one more set of singular cases: when $\bar{r}_1$ is zero then $r_{1}$ is zero for all $Q$ and the series expansions of the other roots changes. For $\bar{r}_{2}$, $\bar{r}_{3}$ and $\bar{r}_{4}$ distinct these are,
\begin{equation}
r_{j} = \bar{r}_{j} + \frac{(18)^{2} }{\prod_{k \ne j} (\bar{r}_{j}-\bar{r}_{k})} Q^{2}+ \frac{(18)^{4} \sum_{k \ne j} (\bar{r}_{k} -\bar{r}_{j})}{\prod_{k \ne j} (\bar{r}_{j}-\bar{r}_{k})^{3}} Q^{4}+ \cdots \qquad (j=2,3,4).
\end{equation}
This series is well behaved when $\bar{r}_{2}$ is zero, \textit{i.e.} when $\bar{r}_{1}=\bar{r}_{2}$, but singular when $\bar{r}_{3}=\bar{r}_{4}=\tilde{r}$ and so we have one final case when $r_1=0$ and the larger roots coalesce:
\begin{align}
r_{2} &= \bar{r}_{2} + \frac{(18)^{2}}{(\tilde{r}-\bar{r}_{2})^{2}} Q^{2} + \frac{2 (18)^{4}}{(\tilde{r}-\bar{r}_{2})^{5}} Q^{4}+ \cdots,\\
r_{k} &= \bar{r}_{k}\pm \frac{18}{\sqrt{\tilde{r}-\bar{r}_{2}}} Q-\frac{(18)^{2}}{2(\tilde{r}-\bar{r}_{2})^{2}} Q^{2} \pm \frac{5 (18)^{3}}{8 (\tilde{r}-\bar{r}_{2})^{7/2}} Q^{3} - \frac{(18)^{4}}{(\tilde{r}-\bar{r}_{2})^{5}} Q^{4} \cdots \qquad (k=3,4),
\end{align}
with the lower signs for $r_{3}$ and the upper signs for $r_{4}$. 

\section{Tables}
\label{app:tables}

The tables below compare the Bohr-Sommerfeld and loop gravity spectra systematically, beginning with the smallest allowed spins and considering all allowed spins up to $(j_1, j_2, j_3, j_4)=(\frac{1}{2},\frac{1}{2}, 3,3)$. To illustrate the rapid improvement of the Bohr-Sommerfeld approximation, as well as the Regge symmetries, two higher spin cases are also included. \\

\begin{tabular}{cccc}
\toprule
\multicolumn{4}{c}{Table} \\
\midrule
$j_1 j_2 j_3 j_4$ & Loop gravity  & Bohr-& Accuracy \\
& & Sommerfeld& \\
\midrule
\multirow{1}{*}{$\frac{1}{2}\ \frac{1}{2}\ \frac{1}{2}\ \frac{1}{2}$}&0.310& 0.252& 19\%\\
\midrule
\multirow{1}{*}{$\frac{1}{2}\ \frac{1}{2}\ \frac{1}{2}\ \frac{3}{2}$}&0& 0& \text{exact}\\
\midrule
\multirow{1}{*}{$\frac{1}{2}\ \frac{1}{2}\ 1\ 1$}&0.396&0.344&13\% \\
\midrule
\multirow{1}{*}{$\frac{1}{2}\ \frac{1}{2}\ 1\ 2$}&0&0& \text{exact} \\
\midrule
\multirow{1}{*}{$\frac{1}{2}\ \frac{1}{2}\ \frac{3}{2}\ \frac{3}{2}$}&0.463& 0.406& 12\% \\
\midrule
\multirow{1}{*}{$\frac{1}{2}\ 1\ 1\ \frac{3}{2}$}&0.498&0.458& 8\% \\
\midrule
\multirow{2}{*}{1\ 1\ 1\ 1}&0&0&\text{exact}\\
&0.620&0.565&9\% \\
\midrule
\multirow{1}{*}{$\frac{1}{2}\ \frac{1}{2}\ \frac{3}{2}\ \frac{5}{2}$}&0&0& \text{exact} \\
\midrule
\multirow{1}{*}{$\frac{1}{2}\ 1\ 1\ \frac{5}{2}$}&0&0& \text{exact} \\
\midrule
\multirow{1}{*}{$\frac{1}{2}\ \frac{1}{2}\ 2\ 2 $}&0.521&0.458& 12\%\\
\midrule
\multirow{1}{*}{$\frac{1}{2}\ 1\  \frac{3}{2}\ 2$}&0.577&0.535& 7\% \\
\midrule
\multirow{1}{*}{$1\ 1\ 1\ 2$}&0.620&0.597& 4\% \\
\midrule
\multirow{1}{*}{$\frac{1}{2}\ \frac{3}{2}\ \frac{3}{2}\ \frac{3}{2}$}&0.620&0.597& 4\%\\
\midrule
\multirow{2}{*}{$1\ 1\ \frac{3}{2}\ \frac{3}{2}$}&0&0&\text{exact}\\
&0.752&0.706& 6\%\\
\bottomrule
\end{tabular}

\begin{tabular}{cccc}
\toprule
\multicolumn{4}{c}{Table} \\
\midrule
$j_1 j_2 j_3 j_4$ & Loop gravity  & Bohr-& Accuracy \\
& & Sommerfeld& \\
\midrule
\multirow{1}{*}{$\frac{1}{2}\ \frac{1}{2}\ 2\ 3$}&0&0& \text{exact} \\
\midrule
\multirow{1}{*}{$\frac{1}{2}\ 1\ \frac{3}{2}\ 3$}&0&0& \text{exact} \\
\midrule
\multirow{1}{*}{$1\ 1\ 1\ 3$}&0&0& \text{exact} \\
\midrule
\multirow{1}{*}{$ \frac{1}{2}\ \frac{1}{2}\ \frac{5}{2}\ \frac{5}{2}$}&0.573&0.504& 12\% \\
\midrule
\multirow{1}{*}{$ \frac{1}{2}\ 1\ 2\ \frac{5}{2} $}&0.644&0.599& 7\%\\
\midrule
\multirow{1}{*}{$\frac{1}{2}\ \frac{3}{2}\ \frac{3}{2}\ \frac{5}{2}$}&0.664&0.621& 6\% \\
\midrule
\multirow{1}{*}{$ 1\ 1\ \frac{3}{2}\ \frac{5}{2}$}&0.713&0.692& 3\%\\
\midrule
\multirow{1}{*}{$\frac{1}{2}\ \frac{3}{2}\ 2\ 2 $}&0.713&0.692& 3\% \\
\midrule
\multirow{2}{*}{$1\ 1\ 2\ 2$}&0&0&  \text{exact} \\
&0.858&0.812& 5\%\\
\midrule
\multirow{2}{*}{$1\ \frac{3}{2}\ \frac{3}{2}\ 2$}&0&0& \text{exact} \\
&0.903&0.867& 4\% \\
\midrule
\multirow{2}{*}{$\frac{3}{2}\ \frac{3}{2}\ \frac{3}{2}\ \frac{3}{2}$}&0.537&0.452& 16\% \\
&0.992&0.947& 5\%\\
\midrule
\multirow{1}{*}{$\frac{1}{2}\ \frac{1}{2}\ \frac{5}{2}\ \frac{7}{2}$}&0& 0& \text{exact}\\
\midrule
\multirow{1}{*}{$\frac{1}{2}\ 1\ 2\ \frac{7}{2}$}&0& 0& \text{exact}\\
\midrule
\multirow{1}{*}{$\frac{1}{2}\ \frac{3}{2}\ \frac{3}{2}\ \frac{7}{2}$}&0& 0& \text{exact}\\
\midrule
\multirow{1}{*}{$1\ 1\ \frac{3}{2}\ \frac{7}{2}$}&0& 0& \text{exact}\\
\midrule
\multirow{2}{*}{$\frac{1}{2}\ \frac{1}{2}\ 3\ 3$}&0& 0& \text{exact}\\
&0.620&0.546&12\%\\
\midrule
\multicolumn{4}{c}{$\cdots$}\\
\midrule
\multirow{6}{*}{$6\ 6\ 6\ 7$}&1.828&1.795&1.8\%\\
&3.204&3.162&1.3\%\\
&4.225&4.190& 0.8\%\\
&5.133&5.105&0.5\%\\
&5.989&5.967&0.4\%\\
&6.817&6.799&0.3\%\\
\midrule
\multirow{6}{*}{$\frac{11}{2}\ \frac{13}{2}\ \frac{13}{2}\ \frac{13}{2}$}&1.828&1.795&1.8\%\\
&3.204&3.162&1.3\%\\
&4.225&4.190& 0.8\%\\
&5.133&5.105&0.5\%\\
&5.989&5.967&0.4\%\\
&6.817&6.799&0.3\%\\
\bottomrule
\end{tabular}


\begin{thebibliography}{10}

\bibitem{BrackBhaduriBook}
M.~Brack and R.~Bhaduri, {\em Semiclassical Physics}.
\newblock Westview Press, 1997.

\bibitem{Rovelli:1995}
C.~{Rovelli} and L.~{Smolin}, ``{Discreteness of area and volume in quantum
  gravity},'' {\em Nuc. Phys. B} {\bf 442} (1995) 593--619,
  \href{http://arXiv.org/abs/arXiv:gr-qc/9411005}{{\tt arXiv:gr-qc/9411005}}.

\bibitem{Ashtekar:1995}
A.~{Ashtekar} and J.~{Lewandowski}, ``{Differential geometry on the space of
  connections via graphs and projective limits},'' {\em Journal of Geometry and
  Physics} {\bf 17} (Nov., 1995) 191--230,
  \href{http://arXiv.org/abs/arXiv:hep-th/9412073}{{\tt arXiv:hep-th/9412073}}.

A.~{Ashtekar} and J.~{Lewandowski}, ``{Quantum Theory of Geometry II: Volume
  operators},'' {\em Adv. Theor. Math. Phys.} {\bf 1} (1998) 388--429,
  \href{http://arXiv.org/abs/arXiv:gr-qc/9711031}{{\tt arXiv:gr-qc/9711031}}.

\bibitem{Bianchi:2011}
E.~{Bianchi} and H.~M. {Haggard}, ``{Discreteness of the Volume of Space from
  Bohr-Sommerfeld Quantization},'' {\em Physical Review Letters} {\bf 107}
  (2011), no.~1, 011301, \href{http://arXiv.org/abs/1102.5439}{{\tt
  1102.5439}}.

\bibitem{DePietri:1996}
R.~{de Pietri} and C.~{Rovelli}, ``{Geometry eigenvalues and the scalar product
  from recoupling theory in loop quantum gravity},'' {\em Phys. Rev. D} {\bf
  54} (1996) 2664--2690, \href{http://arXiv.org/abs/arXiv:gr-qc/9602023}{{\tt
  arXiv:gr-qc/9602023}}.

\bibitem{Thiemann:1998}
T.~{Thiemann}, ``{Closed formula for the matrix elements of the volume operator
  in canonical quantum gravity},'' {\em J. Math. Phys.} {\bf 39} (1998)
  3347--3371, \href{http://arXiv.org/abs/arXiv:gr-qc/9606091}{{\tt
  arXiv:gr-qc/9606091}}.

\bibitem{Carbone:2002}
G.~{Carbone}, M.~{Carfora}, and A.~{Marzuoli}, ``{Quantum states of elementary
  three-geometry},'' {\em Classical and Quantum Gravity} {\bf 19} (2002)
  3761--3774, \href{http://arXiv.org/abs/arXiv:gr-qc/0112043}{{\tt
  arXiv:gr-qc/0112043}}.

\bibitem{Meissner:2006}
K.~A. {Meissner}, ``{Eigenvalues of the volume operator in loop quantum
  gravity},'' {\em Class. Quant. Grav.} {\bf 23} (2006) 617--625,
  \href{http://arXiv.org/abs/arXiv:gr-qc/0509049}{{\tt arXiv:gr-qc/0509049}}.

\bibitem{Brunnemann:2006}
J.~{Brunnemann} and T.~{Thiemann}, ``{Simplification of the spectral analysis
  of the volume operator in loop quantum gravity},'' {\em Classical and Quantum
  Gravity} {\bf 23} (2006) 1289--1346,
  \href{http://arXiv.org/abs/arXiv:gr-qc/0405060}{{\tt arXiv:gr-qc/0405060}}.

\bibitem{Brunnemann:2008a}
J.~{Brunnemann} and D.~{Rideout}, ``{Properties of the volume operator in loop
  quantum gravity: I. Results},'' {\em Classical and Quantum Gravity} {\bf 25}
  (2008), no.~6, 065001, \href{http://arXiv.org/abs/0706.0469}{{\tt
  0706.0469}}.

\bibitem{Brunnemann:2008b}
J.~{Brunnemann} and D.~{Rideout}, ``{Properties of the volume operator in loop
  quantum gravity: II. Detailed presentation},'' {\em Classical and Quantum
  Gravity} {\bf 25} (2008), no.~6, 065002,
  \href{http://arXiv.org/abs/0706.0382}{{\tt 0706.0382}}.

\bibitem{Rovelli:2004}
C.~Rovelli, {\em Quantum Gravity}.
\newblock Cambridge University Press, 2004.

\bibitem{Thiemann:2007}
T.~Thiemann, {\em Modern Canoncial Quantum General Relativity}.
\newblock Cambridge Univ. Press, 2007.

\bibitem{Ashtekar:2004}
A.~{Ashtekar} and J.~{Lewandowski}, ``{Background independent quantum gravity:
  a status report},'' {\em Classical and Quantum Gravity} {\bf 21} (2004) 53,
  \href{http://arXiv.org/abs/arXiv:gr-qc/0404018}{{\tt arXiv:gr-qc/0404018}}.

\bibitem{Rovelli:2010bf}
C.~Rovelli, ``{Loop quantum gravity: the first twenty five years},'' {\em
  Class.Quant.Grav.} {\bf 28} (2011) 153002,
\href{http://arXiv.org/abs/1012.4707}{{\tt 1012.4707}}.

\bibitem{Rovelli:2011eq}
C.~Rovelli, ``{Zakopane lectures on loop gravity},''
\href{http://arXiv.org/abs/1102.3660}{{\tt 1102.3660}}.

\bibitem{Bianchi:2011a}
E.~{Bianchi}, P.~{Don{\'a}}, and S.~{Speziale}, ``{Polyhedra in loop quantum
  gravity},'' {\em Phys. Rev. D} {\bf 83} (2011), no.~4, 044035,
  \href{http://arXiv.org/abs/1009.3402}{{\tt 1009.3402}}.

\bibitem{Perez:2012wv}
A.~Perez, ``{The Spin Foam Approach to Quantum Gravity},''
  \href{http://arXiv.org/abs/1205.2019}{{\tt 1205.2019}}. to appear in
  \emph{Liv.Rev.Rel.}

\bibitem{Barbieri:1998}
A.~{Barbieri}, ``{Quantum tetrahedra and simplicial spin networks},'' {\em
  Nuclear Physics B} {\bf 518} (1998) 714--728,
  \href{http://arXiv.org/abs/arXiv:gr-qc/9707010}{{\tt arXiv:gr-qc/9707010}}.

\bibitem{Kapovich:1996}
M.~Kapovich and J.~J. Millson, ``The symplectic geometry of polygons in
  Euclidean space,'' {\em J. Diff. Geom.} {\bf 44} (1996) 479--513.

\bibitem{Baez:1999tk}
J.~C. Baez and J.~W. Barrett, ``{The Quantum tetrahedron in three-dimensions
  and four-dimensions},'' {\em Adv.Theor.Math.Phys.} {\bf 3} (1999) 815--850,
\href{http://arXiv.org/abs/gr-qc/9903060}{{\tt gr-qc/9903060}}.

\bibitem{Conrady:2009px}
F.~Conrady and L.~Freidel, ``{Quantum geometry from phase space reduction},''
  {\em J.Math.Phys.} {\bf 50} (2009) 123510,
\href{http://arXiv.org/abs/0902.0351}{{\tt 0902.0351}}.

\bibitem{Major:1999}
S.~A. {Major}, ``{Operators for quantized directions},'' {\em Classical and
  Quantum Gravity} {\bf 16} (1999) 3859--3877,
  \href{http://arXiv.org/abs/arXiv:gr-qc/9905019}{{\tt arXiv:gr-qc/9905019}}.

\bibitem{Fairbairn:2004}
W.~{Fairbairn} and C.~{Rovelli}, ``{Separable Hilbert space in loop quantum
  gravity},'' {\em Journal of Mathematical Physics} {\bf 45} (July, 2004)
  2802--2814, \href{http://arXiv.org/abs/arXiv:gr-qc/0403047}{{\tt
  arXiv:gr-qc/0403047}}.

\bibitem{Grot:1996}
N.~Grot and C.~Rovelli, ``{Moduli-space structure of knots with
  intersections},'' {\em J. Math. Phys.} {\bf 37} (1996) 3014--3021,
  \href{http://arXiv.org/abs/gr-qc/9604010}{{\tt gr-qc/9604010}}.

\bibitem{Chakrabarti:1964}
A.~Chakrabarti, ``On the coupling of 3 angular momenta,'' {\em Ann. Henri
  Poincar\'e} {\bf 1} (1964) 301--327.

\bibitem{LevyLeblond:1965}
J.~L\'evy-Leblond and M.~L\'evy-Nahas, ``Symmetrical coupling of three angular
  momenta,'' {\em J. Math. Phys.} {\bf 6} (1965) 1372--1380.

\bibitem{Aquilanti:2007}
V.~Aquilanti, H.~M. Haggard, R.~G. Littlejohn, and L.~Yu, ``Semiclassical
  analysis of Wigner $3j$-symbol,'' {\em J. Phys. A} {\bf 40} (2007), no.~21,
  5637.

\bibitem{Minkowski:1897}
H.~Minkowski, ``Allgemeine Lehrs\"atze \"uber die konvexe polyeder,'' {\em
  Nachr. Ges. Wiss.} (1897) G\"ottingen 198--219.

\bibitem{Marsden:1999b}
J.~E. Marsden and T.~S. Ratiu, {\em Introduction to Mechanics and Symmetry}.
\newblock Springer, 1999.

\bibitem{Aquilanti:2012}
V.~{Aquilanti}, H.~M. {Haggard}, A.~{Hedeman}, N.~{Jeevanjee}, R.~G.
  {Littlejohn}, and L.~{Yu}, ``{Semiclassical Mechanics of the Wigner
  $6j$-Symbol},'' {\em J. Phys. A:Math. Theor.} {\bf 45} (2012) 065209,
  \href{http://arXiv.org/abs/1009.2811}{{\tt 1009.2811}}.

\bibitem{Regge:1959}
T.~{Regge}, ``{Symmetry Properties of Racah's Coefficients},'' {\em Nuovo Cim.}
  {\bf 11} (1959) 116--117,
  \href{http://arXiv.org/abs/arXiv:math-ph/9812013}{{\tt
  arXiv:math-ph/9812013}}.

\bibitem{Roberts:1999}
J.~{Roberts}, ``{Classical 6j-symbols and the tetrahedron},'' {\em Geom. Top.}
  {\bf 3} (1999) 21--66, \href{http://arXiv.org/abs/arXiv:math-ph/9812013}{{\tt
  arXiv:math-ph/9812013}}.

\bibitem{Littlejohn:2009}
R.~G. Littlejohn and L.~Yu, ``{Uniform semiclassical approximation for the
  Wigner $6j$-symbol in terms of rotation matrices},'' {\em J. Phys. Chem. A.}
  {\bf 113} (2009) 14904--14922.


\bibitem{Girelli:2005}
F.~{Girelli} and E.~R. {Livine}, ``{Reconstructing quantum geometry from
  quantum information: spin networks as harmonic oscillators},'' {\em Clas. and
  Quant. Grav.} {\bf 22} (2005) 3295--3313,
  \href{http://arXiv.org/abs/arXiv:gr-qc/0501075}{{\tt arXiv:gr-qc/0501075}}.

L.~{Freidel} and E.~R. {Livine}, ``{The fine structure of SU(2) intertwiners
  from U(N) representations},'' {\em J. Math. Phys.} {\bf 51} (2010), no.~8,
  082502, \href{http://arXiv.org/abs/0911.3553}{{\tt 0911.3553}}.


\bibitem{Freidel:2010a}
L.~{Freidel} and S.~{Speziale}, ``{Twisted geometries: A geometric
  parametrization of SU(2) phase space},'' {\em Phys. Rev. D} {\bf 82} (2010),
  no.~8, 084040--1--16, \href{http://arXiv.org/abs/1001.2748}{{\tt 1001.2748}}.

L.~{Freidel} and S.~{Speziale}, ``{From twistors to twisted geometries},'' {\em
  Phys. Rev. D} {\bf 82} (2010), no.~8, 084041--1--5,
  \href{http://arXiv.org/abs/1006.0199}{{\tt 1006.0199}}.

\bibitem{Major:2010}
S.~A. {Major}, ``{Shape in an atom of space: exploring quantum geometry
  phenomenology},'' {\em Class. Quant. Grav.} {\bf 27} (2010), no.~22,
  225012--+, \href{http://arXiv.org/abs/1005.5460}{{\tt 1005.5460}}.

\bibitem{Yutsis:1962}
A.~P. Yutsis, I.~B. Levinson, and V.~V. Vanagas
{\em The Theory of Angular Momentum}  
\newblock Jerusalem: Israel Program for Scientific Translations, 1962

\end{thebibliography}

\providecommand{\href}[2]{#2}\begingroup\raggedright\endgroup

\end{document}